\documentclass[fleqn,usenatbib]{mnras}

\usepackage{newtxtext,newtxmath}
\usepackage[T1]{fontenc}

\DeclareRobustCommand{\VAN}[3]{#2}
\let\VANthebibliography\thebibliography
\def\thebibliography{\DeclareRobustCommand{\VAN}[3]{##3}\VANthebibliography}

\usepackage{tikz}
\usepackage{orcidlink}
\usepackage{comment}

\usepackage{float}

\usepackage{graphicx}	

\usepackage{multirow}
\usepackage{xcolor}
\usepackage{manyfoot}
\usepackage{booktabs}
\usepackage{changepage}
\usepackage[percent]{overpic}
\usepackage{xfrac}

\usepackage[normalem]{ulem}

\renewcommand{\arraystretch}{1.1}
\usepackage{booktabs,threeparttable}

\usepackage{subfigure}

\newcommand{\specURL}{\url{https://ora.ox.ac.uk/objects/uuid:5032f338-aff0-4089-9700-03dc5c965113}}

\newcommand{\msun}{\mbox{M$_{\odot}$}}
\newcommand{\kms}{\mbox{$\rm{km}\,s^{-1}$}}

\newcommand{\I}{{\sc i}}
\newcommand{\II}{{\sc ii}}
\newcommand{\III}{{\sc iii}}
\newcommand{\IV}{{\sc iv}}

\newcommand{\eg}{\mbox{e.g.}}
\newcommand{\ie}{\mbox{i.e.}}
\newcommand{\cf}{\mbox{cf.}}

\newcommand{\rpro}{\mbox{$r$-process}}
\newcommand{\Aval}{\mbox{$A$-values}}
\newcommand{\ergs}{\mbox{erg\,s$^{-1}$}}

\newcommand{\Ye}{$Y_e$}

\newcommand{\YeTwoOne}{\mbox{$Y_e - 0.21$a}}
\newcommand{\YeZeroFive}{\mbox{$Y_e - 0.05$a}}

\newcommand{\ATxx}[1]{\mbox{AT\,#1}}
\newcommand{\SNxx}[1]{\mbox{SN\,#1}}
\newcommand{\GRBxx}[1]{\mbox{GRB\,#1}}
\newcommand{\thisGRB}{\GRBxx{230307A}}

\newcommand{\SpeciesX}[3]{\item \mbox{\textbf{[{#2\,\textsc{#3}]:}}}}

\newcommand{\jwst}{\mbox{\textit{JWST}}}

\newcommand{\spitzer}{\mbox{\textit{Spitzer}}}
\newcommand{\xsh}{\mbox{X-shooter}}

\newcommand{\gfo}{\mbox{AT\,2017gfo}}
\newcommand{\vfi}{\mbox{AT\,2023vfi}}

\newcommand{\ditto}{\texttt{"}}

\title[Spectral analysis of \vfi]{Analysis of the \jwst\ spectra of the kilonova \vfi\ accompanying \thisGRB}

\author[J.~H.~Gillanders \& S.~J.~Smartt]{J.~H.~Gillanders\,\orcidlink{0000-0002-8094-6108}\thanks{E-mail: james.gillanders@physics.ox.ac.uk} \& S.~J.~Smartt\,\orcidlink{0000-0002-8229-1731} \\
Astrophysics sub-Department, Department of Physics, University of Oxford, Keble Road, Oxford, OX1 3RH, UK
}

\date{Accepted XXX. Received YYY; in original form ZZZ}

\pubyear{2025}

\begin{document}
\label{firstpage}
\pagerange{\pageref{firstpage}--\pageref{lastpage}}
\maketitle

\begin{abstract}
    Kilonovae are key to advancing our understanding of \rpro\ nucleosynthesis. To date, only two kilonovae have been spectroscopically observed, \gfo\ and \vfi. Here, we present an analysis of the \textit{James Webb Space Telescope} (\jwst) spectra obtained +29 and +61~days post-merger for \vfi\ (the kilonova associated with \thisGRB). After re-reducing and photometrically flux-calibrating the data, we empirically model the observed X-ray to mid-infrared continua with a power law and a blackbody, to replicate the non-thermal afterglow and apparent thermal continuum $\gtrsim 2$\,\micron. We fit Gaussians to the apparent emission features, obtaining line centroids of $20218_{-38}^{+37}$, $21874 \pm 89$ and $44168_{-152}^{+153}$\,\AA, and velocity widths spanning $0.057 - 0.110$\,c. These line centroid constraints facilitated a detailed forbidden line identification search, from which we shortlist a number of \rpro\ species spanning all three \rpro\ peaks. We rule out Ba\,\II\ and Ra\,\II\ as candidates and propose Te\,\I--\III, Er\,\I--\III\ and W\,\III\ as the most promising ions for further investigation, as they plausibly produce multiple emission features from one (W\,\III) or multiple (Te\,\I--\III, Er\,\I--\III) ion stages. We compare to the spectra of \gfo, which also exhibit prominent emission at $\sim 2.1$\,\micron, and conclude that [Te\,\III]~$\lambda 21050$ remains the most plausible cause of the observed $\sim 2.1$\,\micron\ emission in both kilonovae. However, the observed line centroids are not consistent between both objects, and they are significantly offset from [Te\,\III]~$\lambda 21050$. The next strongest [Te\,\III] transition at 29290\,\AA\ is not observed, and we quantify its detectability. Further study is required, with particular emphasis on expanding the available atomic data to enable quantitative non-LTE spectral modelling.
\end{abstract}

\begin{keywords}
    gamma-ray burst: individual: \thisGRB\ --- neutron star mergers --- line: identification --- atomic data
\end{keywords}


\section{Introduction} \label{SEC: Introduction}

The rapid neutron-capture process (\rpro) is responsible for synthesising approximately half of the elements heavier than the iron peak (\ie, those with atomic number, $Z \gtrsim 30$) in the Universe
\citep{Cowan2021}. There are a number of proposed production sites for \rpro\ nucleosynthesis \citep{Arnould2007, Thielemann2011}. The explosive deaths of massive stars have been linked as viable production sites for decades, with core-collapse supernovae \citep[SNe;][]{Takahashi1994, Woosley1994, Qian1996}, magneto-rotational SNe \citep{Symbalisty1985, Winteler2012, Mosta2018}, and collapsars \citep{MacFadyen1999, Cameron2003, Pruet2003, Siegel2019} all having been proposed to support the extreme conditions needed to initiate the \rpro. Compact binary mergers have also been proposed as ideal production sites; \ie, the merger of a neutron star (NS) with a companion NS or a stellar-mass black hole (BH) \citep{Lattimer1974, Symbalisty1982, Eichler1989, Li1998, Freiburghaus1999, Rosswog1999}. In these cases, the radioactive decay from the ensemble of newly synthesised unstable heavy isotopes powers an electromagnetic (EM) transient, dubbed a kilonova \citep[KN;][]{Metzger2010}. To date, the only direct observations we have of \rpro\ nucleosynthesis are associated with KNe. The first spectroscopically observed KN event was discovered in August 2017 \citep[\gfo;][]{MMApaper2017} with the second observed $\sim 5.5$~years later, in March 2023 \citep[\vfi, associated with \thisGRB;][]{Levan2024,Yang2024}.

Spectroscopic observations obtained for \gfo\ presented the first opportunity to use spectral data to unravel the composition of freshly synthesised \rpro\ material. Several works have modelled the spectra of \gfo, with the aim of linking specific spectral features to individual \rpro\ species; some of these identifications are more well-established than others. Specific features in the early spectra have been proposed to be due to Sr \citep{Watson2019, Domoto2021, Gillanders2022_PaperI}, La and Ce \citep{Domoto2022, Gillanders2023_PaperII}, and Y \citep{Sneppen2023_YII}, while Te has been proposed as a source of emission in the later spectra \citep{Hotokezaka2023,Gillanders2023_PaperII}. Alternative, less well-established identifications have been proposed for the early phase spectral features; these include Cs and Te \citep{Smartt2017}, He \citep{Perego2022, Tarumi2023}, and Rb \citep{Pognan2023}.

Emphasis has been placed on the proposed identifications of Sr in the photospheric stage \citep{Watson2019} and Te as a nebular emission line \citep{Hotokezaka2023,Gillanders2023_PaperII}, since these are even-$Z$ elements that correspond to the first and second \rpro\ abundance peaks, respectively. As such, these elements will be abundant across many different \rpro\ scenarios. \cite{Gillanders2022_PaperI} present plausible composition profiles for kilonovae \citep[based on different $Y_e$ nucleosynthetic trajectories extracted from a realistic hydrodynamical simulation of a binary neutron star merger from][]{Bauswein2013, Goriely2011,Goriely2013,Goriely2015}, which identify both Sr and Te as potentially abundant elements. Therefore, observing evidence of these elements seems plausible, at least from a nucleosynthetic standpoint. Atomic structure also plays an important role; for example, small numbers of valence electrons can result in a few intrinsically strong transitions that are capable of imprinting prominent spectral features, despite a relatively small abundance (\eg, Sr\,\II). However, KN ejecta possess rapid expansion velocities ($v_{\rm ej} \gtrsim 0.1$\,c), which leads to broad and blended spectral features, making line identification studies difficult \citep{Shingles2023,Collins2024}. One solution to this problem is to probe longer wavelengths, into the near-infrared (NIR) and mid-infrared (MIR), where the effects of line blending are reduced due to the lower density of transitions at these longer wavelengths. An alternative workaround to remove ambiguity surrounding the presence of a particular species is to identify multiple features produced by the same element. However, none of the above proposed identifications are linked to more than a single feature. 

For the Te identification in \gfo, \cite{Hotokezaka2023} propose a [Te\,\III] transition between two fine-structure levels in the ground configuration (with a rest wavelength, $\lambda_{\rm vac} = 21050$\,\AA), as the cause for an observed emission feature at $\sim 21000$\,\AA\ in the $+7.4 - 10.4$\,d \xsh\ spectra of \gfo. \cite{Gillanders2023_PaperII} empirically modelled all spectral features in the late-time spectra of \gfo, and found that this emission feature is best explained by a superposition of two emission components, with peak wavelengths of 20590 and 21350\,\AA. \cite{Gillanders2023_PaperII} independently proposed the same [Te\,\III] transition as a candidate for one of these components, but show alternative matches to what may be relatively prominent forbidden transitions belonging to other species. The [Te\,\III] $\lambda21050$ has been astrophysically observed in planetary nebulae \citep{Madonna2018}. 

For feature identifications in the spectroscopic observations of \vfi, \cite{Levan2024} propose the emission feature present in the +29 and +61\,d \textit{James Webb Space Telescope} (\jwst) NIRSpec spectra at $\sim 21500$\,\AA\ is caused by the same [Te\,\III] transition as proposed for \gfo. \cite{Gillanders2023arxiv_GRB230307A} present modelling of these same \jwst\ spectra, and identify the same spectral feature. They show that the feature at +29\,d can be fit with either one or two components, with peak wavelengths of $\approx 21200$\,\AA\ (one component) or $\approx 20500$ and 22800\,\AA\ (two components). Prominent emission persists at this wavelength position to the later +61\,d \jwst\ spectrum, and \cite{Gillanders2023arxiv_GRB230307A} estimate a peak wavelength of $\approx 22000$\,\AA. \cite{Gillanders2023arxiv_GRB230307A} propose the [Te\,\III] 21050\,\AA\ transition as a likely contributor to this emission, but also suggest alternative identifications.

In this paper, we empirically model a re-reduced and carefully calibrated  version of the \jwst\ spectra of \vfi. We carry out an extensive analysis of the spectral features, and search for candidate transitions in both the +29 and +61\,d spectra, beyond the [Te\,\III] transition proposed by \cite{Levan2024} and \cite{Gillanders2023arxiv_GRB230307A}. We quantify whether the $\sim 2.1$\,\micron\ emission feature necessitates multiple components, or whether a single component can satisfactorily reproduce the data. For comparison, we also re-analyse the late-time \gfo\ spectra in an identical manner, to enable robust comparison between the common $\sim 2.1$\,\micron\ emission present in both kilonovae. With constraints on feature properties, we then perform a line identification search, to constrain the most likely contributing species and highlight them as the most pertinent subjects for future detailed atomic data studies. 

We present our paper in the following manner. In Section~\ref{SEC: Spectra}, we summarise the reduction steps for the \vfi\ spectroscopic data that we analyse in this work. We also present the \gfo\ spectra that we model here and compare with \vfi. In Section~\ref{SEC: Spectral fitting}, we describe our method of fitting the spectra, and we present our results. Section~\ref{SEC: Line ID search} contains the details of our line identification search, and here we discuss the favoured candidate transitions. We also present comparisons to previous line identification studies, and to \gfo. Finally, we conclude in Section~\ref{SEC: Conclusions}.

Throughout this manuscript, we adopt the convention of expressing flux and luminosity densities in wavelength space; \ie, $F \equiv F_\lambda, L \equiv L_\lambda$.

\section{Spectra} \label{SEC: Spectra}

Here we briefly outline the sources of spectral data we analyse in this work, both for \vfi\ and \gfo.

\subsection{\vfi} \label{SEC: Spectra - AT2023vfi}

For our analysis of \vfi, we model the \jwst\ NIRSpec spectra obtained +28.9 and +61.4~days post-explosion (the explosion time is well-constrained due to the detection of the $\gamma$-ray burst; \citealt{GRB230307A_detection_GCN33405}). We opted to perform our own  extractions of the 1D spectra of \vfi.

\begin{figure*}
    \centering
    \includegraphics[width=0.8\linewidth]{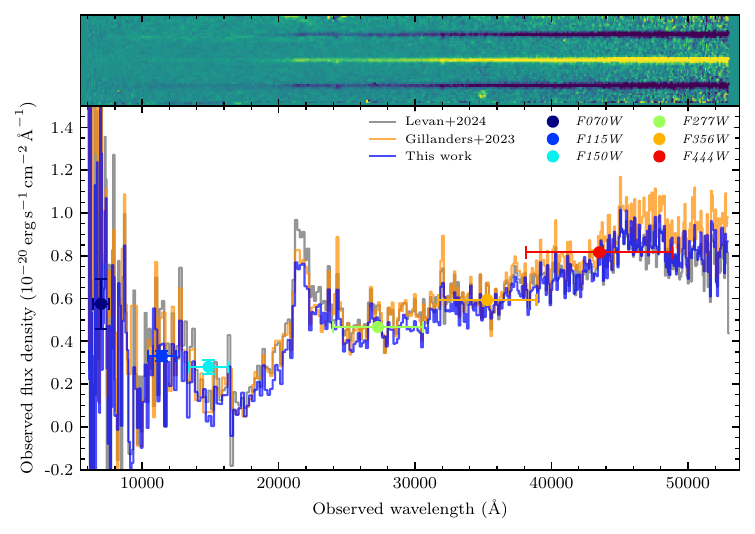}
    \caption{
        \textit{Upper:} +29~day 2D NIRSpec \jwst\ spectrum of \vfi. Note the presence of the three emission features from the background galaxy visible slightly offset from the transient trace, as discussed in the main text.
        \textit{Lower:} Comparison of the different published versions of the +29~day NIRSpec \jwst\ spectrum of \vfi\ (first presented by \protect\citealt{Levan2024} and \protect\citealt{Gillanders2023arxiv_GRB230307A}; grey and orange, respectively) with the version presented in this work (blue). The observed contemporaneous NIRCam photometry from \protect\cite{Levan2024} is overlaid. No extinction or redshift corrections have been applied; all data are plotted as in the observer frame.
    }
    \label{FIG:+29d spectral comparison}
\end{figure*}

\begin{figure*}
    \centering
    \includegraphics[width=0.8\linewidth]{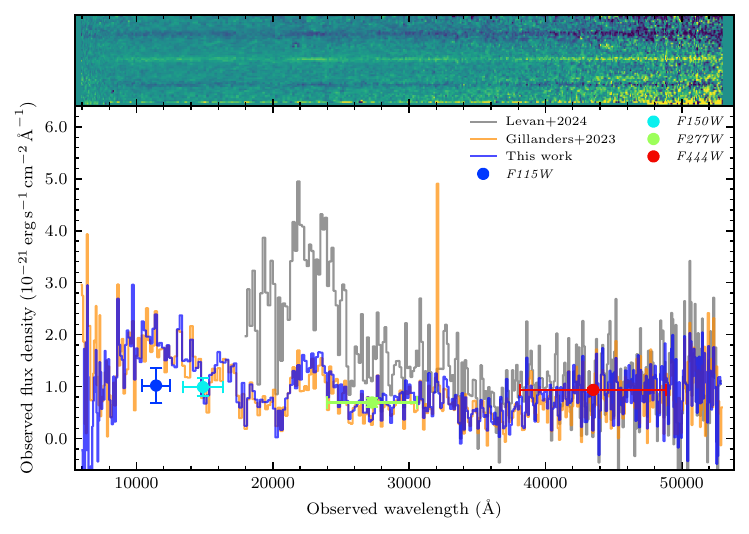}
    \caption{
        Same as Figure~\ref{FIG:+29d spectral comparison}, but for the \jwst\ NIRSpec and NIRCam observations at +61~days.
    }
    \label{FIG:+61d spectral comparison}
\end{figure*}

We first downloaded the level 3 data products derived from the two epochs of NIRSpec observations, available via the MAST webpages.\footnote{\url{https://mast.stsci.edu/portal/Mashup/Clients/Mast/Portal.html}} We note that these level 3 data products were reprocessed on 11th June 2024 with an updated calibration software pipeline\footnote{\mbox{\texttt{cal\_ver\,=\,1.13.3}}} and with updated reference data,\footnote{\mbox{\texttt{CRDS\_CTX\,=\,jwst\_1237.pmap}}} and so should now represent an improvement over the original data products released at the time of observation and those presented by \cite{Levan2024} and \cite{Gillanders2023arxiv_GRB230307A}. These 2D spectra are presented in Figures~\ref{FIG:+29d spectral comparison}~and~\ref{FIG:+61d spectral comparison}.

We extracted the transient spectrum from these following the method outlined by \cite{Horne1986}. We centred an aperture (of width ten pixels) on the transient trace and extracted a 1D spectrum, where the flux values were computed based on the weighted pixel values across the width of the aperture, and the associated error was computed from their variance \citep[see][for further details]{Horne1986}. This method represents the optimal technique for spectral extraction, and combined with the most recent reprocessing, represents the best quality 1D spectra products. The reduced spectra span a wavelength range, $6010 \lesssim \lambda \lesssim 52920$\,\AA.\footnote{This range is slightly smaller than the full wavelength extent of NIRSpec, since we clipped a small amount of extremely noisy data from the blue and red edges.} We note the spectra have a strong dispersion relation between pixel and wavelength scale, with mean values of 105, 65 and 32\,\AA\,pix$^{-1}$ estimated from the central, reddest and bluest ten pixels, respectively. The coarsest dispersion of 200\,\AA\,pix$^{-1}$ is at $\approx 15500$\,\AA.

\begin{figure*}
    \centering
    \subfigure{\includegraphics[width=0.498\textwidth]{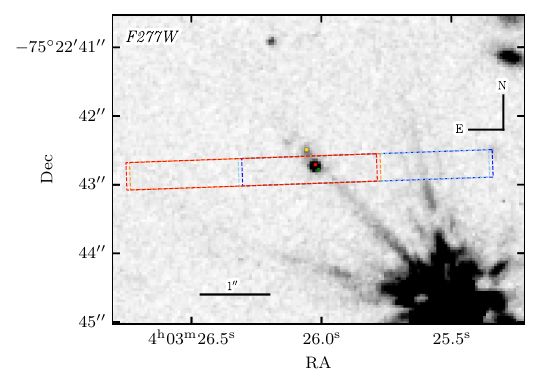}}
    \subfigure{\includegraphics[width=0.498\textwidth]{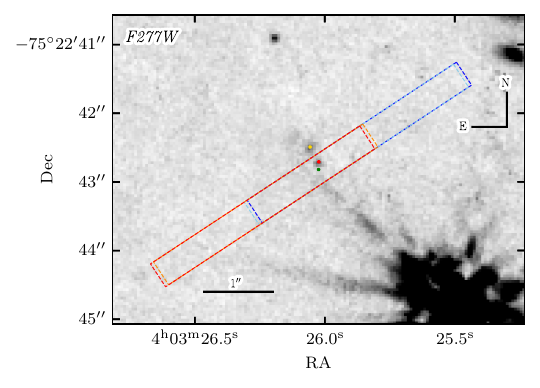}}
    \caption{
        \textit{Left:} +29\,d NIRCam \textit{F277W} image of the field of \vfi/\thisGRB. The location of \vfi\ and the background galaxy are both marked (red and gold dots, respectively). The target location of \vfi\ for this set of observations is marked with a green dot, and corresponds to the central point of the NIRSpec observations. The four dithered NIRSpec slit positions have been overlaid (dashed rectangles). The transient and background galaxy are both clearly visible in the image. The contaminating diffraction spike is also visible.
        \textit{Right:} Same as the left panel, but for the +61\,d NIRCam and NIRspec observations.
    }
    \label{FIG:Slit positions}
\end{figure*}

These Horne-extracted NIRSpec spectra were then flux-calibrated using a linear scaling function to match the contemporaneous NIRCam photometry from \cite{Levan2024}. These photometry estimates have been corrected to account for the contamination from the diffraction spike from a nearby foreground star (which contaminates both the NIRCam and NIRSpec observations). Thus, these photometry values represent the true flux of the transient. Calibrating our 1D NIRSpec spectra to these data therefore allows us to effectively correct for (\ie, remove) the contribution from the diffraction spike.

To visualise the observations and the associated sources of contamination, in Figure~\ref{FIG:Slit positions} we present the \textit{F277W} NIRCam images from +29 and +61\,d, with the slit positions from the NIRSpec observations overlaid. The images are centred on the position of \vfi, with the neighbouring background galaxy and contaminating diffraction spike also visible. The overlaid slit positions represent the position and angle of each of the four NIRSpec sub-exposures. At +29\,d, the slit positions were well-placed, such that \vfi\ was centred close to the centre of the slit in each sub-exposure. The background galaxy was partially sampled at the northern edge of the slit, and we note that the galaxy is offset spatially along the length of the slit, leading to spatially distinct emission in the 2D spectrum (see Figure~\ref{FIG:+29d spectral comparison}).

The background galaxy contamination in the 2D NIRSpec spectrum allows the redshift of this galaxy to be estimated. In our new reduction, we see evidence for three emission lines at 23670, 24317 and 32042\,\AA\ (see Figure~\ref{FIG:+29d spectral comparison}), which, if produced by H$\beta$, [O\,\III]~$\lambda 5006.84$ and H$\alpha$, respectively, correspond to a redshift, $z = 3.87$, in good agreement with that reported by \cite{Levan2024}.

At +61~days, the transient has faded substantially, and the slit angle employed was different. In Figure~\ref{FIG:Slit positions}, we again see that the position of \vfi\ is covered by each of the four dithered NIRSpec exposures. However, the orientation of the slit is now such that the background galaxy was not in the slit in any of the NIRSpec sub-exposures. As a result, we do not see evidence for the prominent galaxy emission lines that were present in the +29\,d observation. The observations are again contaminated by the diffraction spike, and as the transient has faded substantially between +29 and +61~days, the relative contamination is much greater. This leads to a significant excess of flux blueward of $\sim 2$\,\micron, and, following our flux calibration method outlined above, we cannot recover a spectrum that matches the photometric observations across the full NIRSpec $\approx 0.7 - 5.3$\,\micron\ wavelength range (see Figure~\ref{FIG:+61d spectral comparison}).

In the previous version of the +61\,d spectrum, there was a strong single pixel excess centred at $\approx 32090$\,\AA, which initially appeared to be due to H$\alpha$ emission from the background galaxy at $z \approx 3.87$. However, the reprocessed data products do not contain this prominent emission spike. Upon further investigation, we determine it was likely the result of a hot pixel or a cosmic ray artefact that has now been successfully masked out in the new reduction of the data. It was simply coincidental that it lay close to H$\alpha$ at $z = 3.87$.

Here we make our reduced \jwst\ spectra publicly available and accessible to the community (see Data Availability). These spectra are the best available versions of these two \jwst\ NIRSpec observations of \vfi. As such, we suggest these are the versions that should be used for future analysis of this event. In Figures~\ref{FIG:+29d spectral comparison}~and~\ref{FIG:+61d spectral comparison}, we present the reductions of the +29 and +61~day spectra compared with the contemporaneous NIRCam photometry, as well as the previously published reductions of these spectra (from \citealt{Levan2024} and \citealt{Gillanders2023arxiv_GRB230307A}). 

Before modelling the spectra, we need to account for extinction and redshift effects. The extinction along the line of sight to \vfi\ is estimated to be $E(B-V) = 0.0758$~AB~mag \citep{Schlafly2011}, and so we de-redden our \jwst\ spectra by this factor. To shift the observed spectra to the transient's rest frame, we need an accurate redshift. The probable host galaxy of \vfi\ has an estimated redshift, \mbox{$z = 0.0647 \pm 0.0003$} \citep{Gillanders2023GCN.33485_GRB230307A_redshift, Levan2024, Yang2024}. Therefore, we blueshift the spectra by this factor to account for redshift effects.

Finally, for ease of comparison to \gfo, we opted to convert the observed flux in the \jwst\ spectra of \vfi\ to luminosity. The redshift of \vfi\ (\mbox{$z = 0.0647 \pm 0.0003$}) corresponds to a luminosity distance, $D_{\textsc{l}} \simeq 302$\,Mpc \citep[assuming Planck $\Lambda$CDM cosmology with a Hubble constant, \mbox{$H_0 = 67.4$\,km\,s$^{-1}$\,Mpc$^{-1}$}, \mbox{$\Omega_{\textrm{M}} = 0.315$} and \mbox{$\Omega_{\Lambda} = 0.685$};][]{Planck2020}.

\subsection{\gfo} \label{SEC: Spectra - AT2017gfo}

For the spectral analysis of \gfo, we utilise the sequence of \xsh\ spectra obtained by \cite{Pian2017} and \cite{Smartt2017}. Specifically, we use the flux-calibrated spectral data set publicly released by the ENGRAVE collaboration \citep{ENGRAVE2020}. These have been flux-calibrated to photometry, corrected for line-of-sight extinction, and shifted to the rest frame. This spectral data set is readily accessible via the ENGRAVE webpage\footnote{\url{www.engrave-eso.org/AT2017gfo-Data-Release}} (along with release notes detailing the calibration and post-processing steps) and WISeREP\footnote{\url{https://wiserep.weizmann.ac.il}} \citep{wiserep}.

As for \vfi, we convert the spectra of \gfo\ to luminosity, using  $D_{\textsc{l}} \simeq 40.4$\,Mpc \citep[\eg,][]{Hjorth17} for the distance to \gfo.

\section{Spectral fitting} \label{SEC: Spectral fitting}

\subsection{\vfi} \label{SEC: Spectral fitting - AT2023vfi}

To model the \jwst\ spectra of \vfi, we need to account for the different components contributing to the observed flux. There is evidence of contribution from a non-thermal afterglow component, a thermal component, and two broad emission-like features in excess of the continuum (centred at $\sim 2.1$ and 4.4\,\micron). The high-energy observations of the non-thermal afterglow component can be readily fit with a single-component power law at the phases of the \jwst\ observations (see \citealt{Yang2024} for evidence of a jet break at earlier times). This component extends to longer wavelengths, but in the wavelength range of the \jwst\ NIRSpec spectra, it is expected to be subdominant \citep{Gillanders2023arxiv_GRB230307A, Yang2024}. Above $\sim 2$\,\micron, the spectrum of \vfi\ exhibits a rising red continuum, resemblant of a blackbody profile. Here we perform an empirical fit to the observed spectra, accounting for the contribution from all components; \ie, the non-thermal afterglow (AG), the thermal blackbody (BB), and the emission-like features (approximated as Gaussians), using:
\begin{equation}
    L = L_{\textsc{ag}} + L_{\textsc{bb}} + \sum\limits_{i}^{} \Big( L_{\textsc{g}} \Big),
\end{equation}
where $L$ is the total observed luminosity density, and $L_{\textsc{ag}}$, $L_{\textsc{bb}}$, and $L_{\textsc{g}}$
are the observed luminosity densities from the AG, BB, and Gaussian emission components.
These contributions are computed using:
\begin{equation}
    L_{\textsc{ag}} = \alpha \cdot \lambda^{\beta},
\end{equation} \vspace{-2em}
\begin{equation}
    L_{\textsc{bb}} = \pi B(\lambda, T_{\rm ph}) \cdot 4 \pi R_{\rm ph}^{2},
\end{equation} \vspace{-1em}
\begin{equation}
    L_{\textsc{g}} = a \cdot {\rm exp}\left[ - \frac{ (\lambda - b)^{2}}{2c^{2}} \right],
\end{equation}
where $\lambda$ is wavelength, $\alpha$ and $\beta$ are the AG amplitude and exponent, respectively, $B(\lambda, T_{\rm ph})$ is the Planck function, given by:
\begin{equation}
    B(\lambda, T_{\rm ph}) = \frac{2 h c^{2}}{\lambda^5} \left( {\rm exp} \left[ \frac{h c}{\lambda k_{\textsc{b}} T_{\rm ph}} \right] -1 \right)^{-1},
\end{equation}
$T_{\rm ph}$ and $R_{\rm ph}$ are the photospheric temperature and radius, and $a$, $b$ and $c$ are the height, centroid and standard deviation (\ie, `width') of the Gaussian profile, with $i$ representing the number of Gaussian profiles needed to reproduce the observed emission-like features.

We perform an empirical fit to the observed data using Markov Chain Monte Carlo (MCMC) techniques, to deduce the best-fitting parameters for our simple model described above. To best constrain the contribution of the sub-dominant (at least in the wavelength range of NIRCam and NIRSpec) afterglow, we fit the data in two stages. First, we fit the NIRCam photometry jointly with inter/extra-polated X-ray data, to estimate the continuum (\ie, $L_{\textsc{ag}}$ and $L_{\textsc{bb}}$) properties. \cite{Yang2024} present \textit{XMM-Newton} and \textit{Chandra} data for \thisGRB, up to $+37$~days post-explosion. Here we interpolate (for +29\,d) and extrapolate (for +61\,d) these X-ray data to estimate the high-energy contribution of the AG at the phases of the \jwst\ observations. Although there is some uncertainty associated with our inter/extra-polations, these data provide the best possible constraints for the non-thermal afterglow at the phases of the \jwst\ observations.

Then, we use these best-fitting parameters as constraints for the fitting of the NIRSpec data, which allows us to also extract the best-fitting parameters of the Gaussian components, in addition to improving the robustness of the estimated BB properties. We jointly fit the two epochs of observations together, as there are multiple parameters that can be fixed to reduce the degeneracy in our modelling; these include the exponent of the afterglow ($\beta$), and the centroids and `widths' of the Gaussian profiles (\ie, we only allow the feature amplitudes to vary between +29 and +61~days).

Finally, before fitting the data we perform two `data cleaning' steps, to better enable reliable fitting of the NIRSpec data. First, we mask the prominent background galaxy emission in the +29\,d spectrum (see Section~\ref{SEC: Spectra - AT2023vfi}). Second, the \jwst\ observations are contaminated by a diffraction spike from a nearby foreground star (see Section~\ref{SEC: Spectra - AT2023vfi} and Figure~\ref{FIG:Slit positions}). As we were unable to correct for this contamination at shorter wavelengths in the +61\,d spectrum, we truncate the data with $\lambda < 18000$\,\AA\ \citep[observer frame, as in][]{Levan2024}.

\subsubsection{+29~day \jwst\ observation} \label{SEC: Spectral fitting - AT2023vfi - +29 day JWST}

\begin{figure*}
    \centering
    \includegraphics[width=0.8\linewidth]{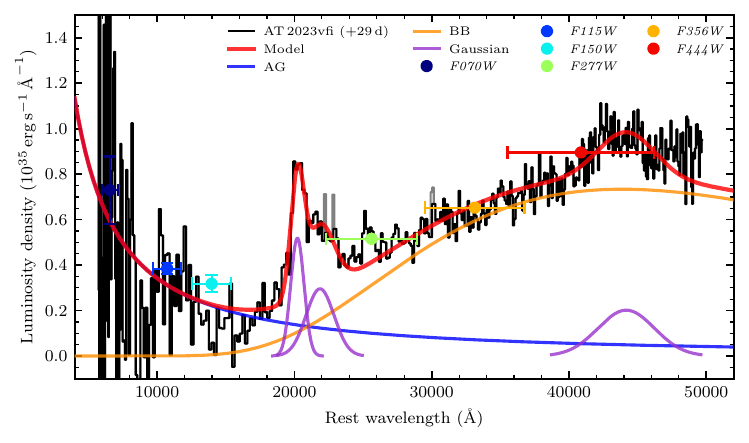}
    \caption{
        +29~day \jwst\ NIRSpec spectrum of \vfi\ (black), compared to our best-fitting model (red). The background galaxy emission lines have been masked. The individual continuum components (\ie, AG, BB) are plotted (blue and orange, respectively), as are the individual Gaussian components (purple). The contemporaneous NIRCam photometry is overlaid. Both the NIRCam and NIRSpec data have been corrected for line-of-sight extinction, and shifted to the rest frame (see Section~\ref{SEC: Spectra - AT2023vfi} for details).
    }
    \label{FIG:AT2023vfi +29d spectral fit (NIRSpec)}
\end{figure*}

Our best-fitting model for the +29~day NIRSpec spectrum is presented in Figure~\ref{FIG:AT2023vfi +29d spectral fit (NIRSpec)}, and the associated best-fitting parameters are summarised in Tables~\ref{TAB:Continuum parameters}~and~\ref{TAB:Feature parameters}. For completeness, we present our best-fitting model extended to include the X-ray data that were used to constrain the AG component in Figure~\ref{FIG:Appendix - +29d spectral fit}.

We find good agreement to the X-ray photometry and NIRSpec spectrum by invoking an afterglow amplitude, \mbox{$\alpha = 6.0_{-1.4}^{+2.0} \times 10^{39}$\,erg\,s$^{-1}$\,\AA$^{-(\beta + 1)}$} and exponent, \mbox{$\beta = -1.31 \pm 0.03$}, and a blackbody with photospheric temperature and radius, $T_{\rm ph} = 660 \pm 10$\,K and \mbox{$R_{\rm ph} = 6.01 \pm 0.14 \times 10^{15}$\,cm}. These values are generally in line with those presented by previous works that fit the NIRSpec data \citep{Gillanders2023arxiv_GRB230307A, Levan2024}. Accurately accounting for the contribution of the continuum enables study of the spectral features that cause deviations from this $\sim$smooth model continuum. Visually, there appear to be three broad emission-like features; a blend of two features centred at $\sim 2.1$\,\micron, and a third centred at $\sim 4.4$\,\micron\ \citep[first identified by][]{Levan2024}.

We find the emission feature at $\sim 2.1$\,\micron\ is best reproduced by the superposition of two distinct Gaussians, with central wavelengths of $20218_{-38}^{+37}$ and $21874 \pm 89$\,\AA. \cite{Levan2024} invoke two components to reproduce this $\sim 2.1$\,\micron\ feature; they report line centroids of $20285 \pm 10$ and $22062 \pm 10$\,\AA, both with full-width, half-maximum (FWHM) velocities, $v_{\textsc{fwhm}} = 0.064$\,c. However, a single emission feature is then invoked when attributing it to [Te\,\III] (see Section~\ref{SEC: Line ID search - Discussion - Te}). \cite{Gillanders2023arxiv_GRB230307A} also fit this feature, and present results obtained from invoking one or two components. For their two-component model, they report feature centroids, $\lambda \approx 20500$, 22800\,\AA, and widths, $v_{\textsc{fwhm}} \approx 0.079$, $0.064$\,c. The variation between the results of \cite{Levan2024} and \cite{Gillanders2023arxiv_GRB230307A} can be interpreted as being due to the effects of different normalisation techniques. Both works fit the continuum slightly differently, which after subtraction leads to some deviation in the perceived feature centroids. The results from our analysis in this work somewhat agree with these previous estimates, but here we perform a (formally) more robust approach of simultaneously fitting the continuum contribution while also fitting the emission features, instead of subtracting the continuum contribution before feature fitting \citep[as in][]{Gillanders2023arxiv_GRB230307A, Levan2024}. The latter approach may not accurately account for the uncertainty in the continuum when obtaining the best-fitting Gaussian parameters.

The feature at $\sim 4.4$\,\micron\ can be well-reproduced with a single Gaussian component, with a centroid, $\lambda = 44168_{-152}^{+153}$\,\AA, and $v_{\textsc{fwhm}} = 0.110_{- 0.003}^{+0.001}$\,c. \cite{Levan2024} do not perform any fitting of this feature. \cite{Gillanders2023arxiv_GRB230307A} report a line centroid, \mbox{$\lambda \approx 43900$\,\AA}, and width, $v_{\textsc{fwhm}} \approx 0.093$\,c. Our result presented here is somewhat deviant to this previously reported value, likely for the reason outlined above.

\subsubsection{+61~day \jwst\ observation} \label{SEC: Spectral fitting - AT2023vfi - +61 day JWST}

\begin{figure*}
    \centering
    \includegraphics[width=0.8\linewidth]{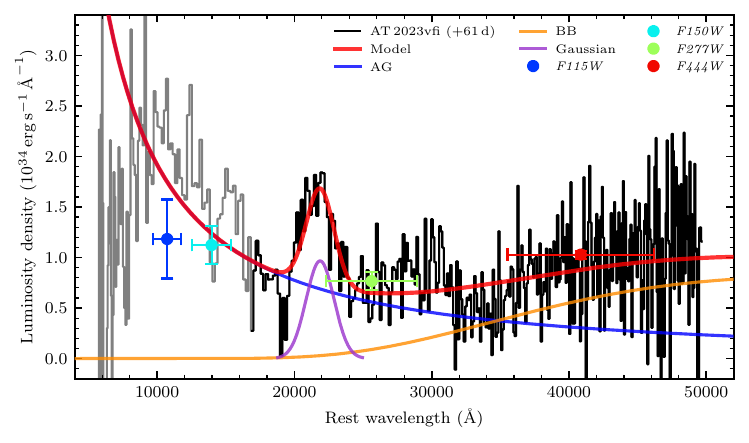}
    \caption{
        Same as Figure~\ref{FIG:AT2023vfi +29d spectral fit (NIRSpec)}, but for the +61~day \jwst\ NIRSpec spectrum of \vfi. Here the contaminated blue end of the spectrum has been masked.
    }
    \label{FIG:AT2023vfi +61d spectral fit (NIRSpec)}
\end{figure*}

\begin{table*}
    \renewcommand*{\arraystretch}{1.3}
    \centering
    \caption{
        Best-fitting continuum parameters for the \jwst\ NIRSpec spectra of \vfi, and for the +10.4\,d \xsh\ spectrum of \gfo.
    }
    \begin{threeparttable}
        \centering
        \begin{tabular}{lccccc}
            \toprule

            Phase (days)       &$\alpha$ ($10^{39}$\,erg\,s$^{-1}$\,\AA$^{-(\beta + 1)}$)        &$\beta$       &$T_{\rm ph}$ (K)       &$R_{\rm ph}$ ($10^{15}$\,cm)      &$v_{\rm ph}$ (c)           \\
            
            \midrule
            \multicolumn{6}{l}{\vfi} \\
            \midrule
    
            +29       &$6.0_{- 1.4}^{+ 2.0}$      &$-1.31 \pm 0.03$    &$660 \pm 10$                &$6.01 \pm 0.14$              &$0.080 \pm 0.002$            \\
            +61       &$3.4_{- 0.8}^{+ 1.1}$      &\ditto              &$500_{- 20}^{+ 30}$         &$3.95_{- 0.61}^{+ 0.62}$     &$0.025 \pm 0.004$            \\
    
            \midrule        
            \multicolumn{6}{l}{\gfo} \\
            \midrule
            
            +10.4     &$-$                        &$-$                 &$1630_{- 110}^{+ 150}$      &$2.09_{- 0.43}^{+ 0.42}$     &$0.077_{- 0.016}^{+ 0.015}$  \\
    
            \bottomrule
        \end{tabular}
        \begin{tablenotes}
            \small
            \item[] \textbf{Note.} Errors are quoted to $1 \sigma$ significance.
        \end{tablenotes}
    \end{threeparttable}
    \label{TAB:Continuum parameters}
\end{table*}

\begin{table*}
    \renewcommand*{\arraystretch}{1.2}
    \centering
    \caption{
        Best-fitting parameters for the emission features in the \jwst\ NIRSpec spectra of \vfi, and for the +10.4\,d \xsh\ spectrum of \gfo.
    }
    \begin{threeparttable}
        \centering
        \begin{tabular}{lccccc}
            \toprule
            
            Phase       &Approx.\ $\lambda_{\rm peak}$       &$\lambda_{\rm peak}$                  &$v_{\textsc{fwhm}}$             &Peak                                       &Luminosity                    \\
            (days)      &(\micron)                           &(\AA)                                 &(c)                             &(10$^{34}$\,erg\,s$^{-1}$\,\AA$^{-1}$)     &(10$^{37}$\,erg\,s$^{-1}$)    \\
            \midrule

            \multicolumn{6}{l}{\vfi} \\
            \midrule
            
            +29         &2.0                                 &$20218_{- 38}^{+ 37}$                 &$0.057_{- 0.005}^{+ 0.006}$     &$5.17_{- 0.34}^{+ 0.35}$                   &$\approx 6.4$                 \\
            +29         &2.2                                 &$21874 \pm 89$                        &$0.109_{- 0.004}^{+ 0.002}$     &$2.96 \pm 0.20$                            &$\approx 7.6$                 \\
            +29         &4.4                                 &$44168_{- 152}^{+ 153}$               &$0.110_{- 0.003}^{+ 0.001}$     &$2.01 \pm 0.14$                            &$\approx 10$                  \\

            +61         &2.2                                 &$21874 \pm 89$                        &$0.109_{- 0.004}^{+ 0.002}$     &$0.97 \pm 0.09$                            &$\approx 2.5$                 \\

            \midrule
            \multicolumn{6}{l}{\gfo} \\
            \midrule

            +10.4       &2.05                                &$20530_{- 190}^{+ 140}$\tnote{*}       &$0.130 \pm 0.011$\tnote{*}     &$34 \pm 19$                                &$\approx 100$                 \\
            +10.4       &2.13                                &$21300_{- 260}^{+ 330}$\tnote{*}       &\ditto                         &$62 \pm 17$                                &$\approx 180$                 \\
            
            \bottomrule
        \end{tabular}        
        \begin{tablenotes}
            \small
            \item[] \textbf{Note.} Errors are quoted to $1 \sigma$ significance.
            \item[*] Our fitting procedure invokes fixed centroids and $v_{\textsc{fwhm}}$ values across the $+7.4 - 10.4$\,d sequence of \xsh\ spectra of \gfo. Thus, the $\lambda_{\rm peak}$ and $v_{\textsc{fwhm}}$ values quoted here for the +10.4\,d spectrum also represent the best-fitting values across the +7.4, +8.4 and +9.4\,d spectra.
        \end{tablenotes}
    \end{threeparttable}
    \label{TAB:Feature parameters}
\end{table*}

Performing a similar analysis on the +61\,d spectrum proves more difficult, since the transient has faded significantly, causing a drop in the signal-to-noise ratio (SNR). Additionally, it is no longer clear whether the spectrum can be accurately represented by a thermal continuum. Nevertheless, we apply the same method and recover a reasonable match to the data. We present our best-fitting model in Figures~\ref{FIG:AT2023vfi +61d spectral fit (NIRSpec)}~and~\ref{FIG:Appendix - +61d spectral fit}, and associated model parameters in Tables~\ref{TAB:Continuum parameters}~and~\ref{TAB:Feature parameters}.

Our best-fitting model invokes an afterglow amplitude, \mbox{$\alpha = 3.4_{-0.8}^{+1.1} \times 10^{39}$\,erg\,s$^{-1}$\,\AA$^{-(\beta + 1)}$}, with the same exponent as for +29\,d (\mbox{$\beta = -1.31 \pm 0.03$}), and a blackbody with photospheric temperature and radius, $T_{\rm ph} = 500_{-20}^{+30}$\,K and $R_{\rm ph} = 3.95_{-0.61}^{+0.62} \times 10^{15}$\,cm. The decrease in $\alpha$ is expected, and is consistent with a fading source. We find no reason to invoke a variation in $\beta$,\footnote{Performing an independent fitting process to the +29 and +61\,d spectra recovers matching $\beta$ values.} indicating there is no evolution in the AG slope between +29 and +61~days \citep[see also][]{Yang2024}.

Visually, there is still a clear flux excess at $\sim 2.1$\,\micron, the location of prominent emission at +29~days. There does not appear to be any similar excess at $\sim 4.4$\,\micron, the location of the other prominent flux excess at +29~days.

We find that we can adequately match this emission component by invoking the same centroid and width as was needed to reproduce the redder component of the blended $\sim 2.1$\,\micron\ feature at +29~days (\ie, $\lambda = 21874 \pm 89$\,\AA\ and $v_{\textsc{fwhm}} = 0.109_{-0.004}^{+0.002}$\,c). \cite{Levan2024} do not analyse this feature at +61~days. \cite{Gillanders2023arxiv_GRB230307A} fit it as a single component, and retrieve a line centroid, $\lambda \approx 22000$\,\AA, and width, $v_{\textsc{fwhm}} \approx 0.173$\,c. The mismatch between this previously reported value and our result here is again likely down to the different approaches for how the continuum was treated (see Section~\ref{SEC: Spectral fitting - AT2023vfi - +29 day JWST}).

Next, we look at -- and compare to -- the prominent emission features in \gfo. For consistency, we need to first analyse them in a similar manner.

\subsection{\gfo} \label{SEC: Spectral fitting - AT2017gfo}

As highlighted in Section~\ref{SEC: Introduction}, there is a notable emission feature present at $\sim 2.1$\,\micron\ in the spectra of both \vfi\ and \gfo. This is easily visually inferred by comparing the spectra of both transients (see \eg, Figure~\ref{FIG:ATs 2017gfo + 2023vfi line IDs}). To quantify the agreement between this feature in these two different transient events, we perform the same analysis on the late-time \xsh\ spectra of \gfo\ as we performed above for the two \jwst\ spectra of \vfi\ (as described in Section~\ref{SEC: Spectral fitting - AT2023vfi}), with one exception. The viewing angle of \gfo\ was sufficiently off-axis \citep[$\lesssim 30 ^\circ$;][]{2019NatAs...3..940H,Dhawan2020,Wang2023} for the contribution of the afterglow contribution at the times and wavelengths sampled by \xsh\ to be negligible, and so we do not include an afterglow component in our fitting routines (\ie, we fit the data considering $L_{\textsc{bb}}$ and $L_{\textsc{g}}$ contributions only).

The late-phase spectra of \gfo\ ($> 7$~days) are not easily reproduced by a single-temperature blackbody profile \citep[see \eg,][]{Waxman2018, Gillanders2022_PaperI}. The near-ultraviolet (near-UV) and optical wavelengths (\mbox{$\lambda < 7000$\,\AA}) exhibit blanketed flux suppression, while there are multiple prominent spectral features across the \mbox{$\sim 0.3 - 2.5$\,\micron} window of \xsh\ \citep[see \eg,][]{Gillanders2022_PaperI, Gillanders2023_PaperII}. However, the NIR regime above $\sim 1.5$\,\micron\ for the $+7.4 - 10.4$~day \xsh\ spectra can be well-reproduced with a single-temperature blackbody and three emission features centred at $\sim 1.6$, 2.05 and 2.13\,\micron. \cite{Gillanders2023_PaperII} show that the redward-shifting nature of the emission at $\sim 2.1$\,\micron\ in the late-time spectra of \gfo\ (observed feature peak shifts from 2.06\,\micron\ at +7.4\,d, to 2.10\,\micron\ at +10.4\,d) is best reproduced by invoking two blended components, with evolving relative strengths to explain the peak wavelength shift. Here we also find that the data are best matched by invoking two components for this emission feature (see below). Since we are interested specifically in the properties of the emission around $\sim 2.1$\,\micron\ for comparison to \vfi, we focus our efforts on only modelling the NIR regime ($\lambda = 1.5 - 2.5$\,\micron) in this fashion, and ignore the complex continuum and spectral features at shorter wavelengths.

We stress that the continuum $T$ and $R$ values invoked in this analysis are not meant to be interpreted as estimates for the photospheric temperature and radius; the values obtained are merely those that were estimated through our simple empirical modelling to reproduce the `local continuum' of the two NIR emission features from $1.5 - 2.5$\,\micron. These values should not be used to infer the photospheric properties across the full \xsh-sampled wavelength range (if a $\nu$-independent photospheric approximation is even a valid assumption at these late phases).

\subsubsection{$\mathit{+7.4 - 10.4}$~day \xsh\ observations}

Here we fit the $+7.4 - 10.4$~day \xsh\ spectra with fixed feature centroids and widths, while the feature peak and continuum values are allowed to freely vary between epochs. From this approach, we obtain reasonable fits for the $1.5 - 2.5$\,\micron\ spectral range for all four epochs (see Figure~\ref{FIG:AT2017gfo spectral fits}). The best-fitting parameters invoked for our fitting of the +10.4\,d spectrum are presented in Tables~\ref{TAB:Continuum parameters}~and~\ref{TAB:Feature parameters}.

With our local continua reasonably well parameterised, we now focus on the properties of the emission features. While we have fit the $\sim 1.6$\,\micron\ emission component to accurately constrain the NIR \xsh\ data, we are not specifically interested in the results obtained for this feature. Nevertheless, for posterity and for comparison with the work of \cite{Gillanders2023_PaperII}, we present them here. For the $\sim 1.6$\,\micron\ feature, we find best-fitting values for the centroid and velocity width of $\lambda_{\rm peak} = 15860 \pm 90$\,\AA\ and \mbox{$v_{\textsc{fwhm}} = 0.080^{+0.028}_{-0.018}$\,c}, respectively. \cite{Gillanders2023_PaperII} estimate $\lambda_{\rm peak} \approx 15800$\,\AA\ and $v_{\textsc{fwhm}} \approx 0.097$\,c, which agree well with our new estimates. For the blended $\sim 2.1$\,\micron\ emission feature, \cite{Gillanders2023_PaperII} estimate $\lambda_{\rm peak} \approx 20590$, 21350\,\AA\ with $v_{\textsc{fwhm}} \approx 0.130$\,c for both components. Again, these values match our new estimates well ($\lambda_{\rm peak} = 20530^{+140}_{-190}$, $21300^{+330}_{-260}$\,\AA\ and $v_{\textsc{fwhm}} = 0.130 \pm 0.011$\,c for both components). We note that in \cite{Gillanders2023_PaperII}, they assumed a flat local continuum, normalised the spectrum, and then modelled each emission feature with a Gaussian profile (one or multiple). Here we perform a more complete analysis, whereby we simultaneously model the continuum and emission features. Despite the different treatments for the local continua, we obtain similar results for the properties of the emission features.

Our models here reproduce the data well, as shown by the comparisons presented in Figure~\ref{FIG:AT2017gfo spectral fits}. We conclude that a single transition at 1.5860\,\micron, and two blended transitions at 2.0530 and 2.1300\,\micron\ can explain the spectra well, but the velocity width of the lines are discrepant, with $v_{\textsc{fwhm}} ({\rm c}) = 0.080^{+0.028}_{-0.018}$ for the 1.5860\,\micron\ line and $v_{\textsc{fwhm}} ({\rm c}) = 0.130 \pm 0.011$ for the 2.0530, 2.1300\,\micron\ lines (see Table~\ref{TAB:Feature parameters}). This may indicate that they arise from different regions of the ejecta. Alternatively, it could be evidence for these features being composed of a blend of multiple/many emission lines, in which case we cannot accurately constrain the velocity of the line-forming region via this fitting procedure.

\section{Line identification search} \label{SEC: Line ID search}

With accurate estimates for the positions of the feature centroids in the \jwst\ spectra of \vfi\ (and the late-time \xsh\ spectra of \gfo), we now perform a detailed line identification study to search for candidate transitions that plausibly produce these emission features. Here we focus specifically on forbidden transitions, which are expected to dominate late-time emission features in kilonova spectra.

Forbidden transitions have already been shown to be important for understanding the nebular spectral features of core-collapse and type~Ia SNe \citep[see \eg,][]{Mazzali2001,Dessart2011, Jerkstrand2017-HB, Shingles2020}, and several works have also highlighted their importance and the potential effects they may have on late-phase KN spectra \citep[\eg,][]{Hotokezaka2021, Hotokezaka2022, Pognan2022, Pognan2022_nlte, Pognan2023}. However, at least for the KN case, detailed forbidden transition studies are hampered by the incompleteness of the line lists for \rpro\ species. Therefore, recent studies have resorted to using forbidden line information derived from theoretical calculations \citep{Hotokezaka2021, Pognan2023}. These typically have uncertain wavelengths, which impacts line identification studies.

\cite{Gillanders2023_PaperII} follow a different approach to search for candidate forbidden transitions that may be responsible for producing the emission features in the late-phase spectra of the kilonova \gfo. They used the level information from the National Institute of Standards and Technology Atomic Spectra Data base \citep[NIST ASD;][]{NIST2023} to estimate the transition wavelengths for all possible transitions between all levels of each ion. Then, by considering the transition rules for forbidden transitions, a shortlist was obtained for each ion under investigation. Since all energy levels within the NIST ASD have been critically evaluated, they possess accurate energies, meaning the transitions derived via this method have robust wavelengths, accurate enough for line identification studies. However, this method does suffer from the issue of not providing oscillator strengths and collision strengths (in addition to it likely being incomplete for the heaviest elements). Such information is needed to undertake a quantitative study of KNe including non-local thermodynamic equilibrium (non-LTE) effects, such as that carried out by \cite{Pognan2023}.

Throughout this section, unless otherwise highlighted, we quote the wavelengths of our candidate transitions as in vacuum; to access the equivalent air wavelengths, see Table~\ref{TAB:Candidate lines}.

\subsection{Methodology} \label{SEC: Line ID search - Methodology}

Here we follow the method presented by \cite{Gillanders2023_PaperII}. Below we summarise the main steps, and highlight any deviations from the previously established methodology.

We begin by extracting the level information from the NIST ASD \citep{NIST2023} for all elements with $Z = 30 - 92$ (Zn to U), for the lowest four ionisation stages (neutral to triply ionised; \ie, $\textsc{i} - \textsc{iv}$). Previous analysis focussed on $Z = 38 - 92$ (Sr to U), and on the lowest three ion stages (neutral to doubly ionised; motivated by LTE ionisation arguments). Here we extend our analysis to lower $Z$ (for completeness and to capture line information for the lightest species; see Figure~\ref{FIG:r-process abundances}) and we also consider the triply ionised stage. While we do not anticipate significant quantities of ejecta material to be triply ionised due to thermal ionisation effects, it is reasonable to expect this level of ionisation can be achieved through non-thermal mechanisms \citep[see][]{Pognan2023}.

With this level information, we compute the wavelengths of all possible transitions between all levels for each ion. We truncate the list to exclude any lines that originate from an upper level with energy greater than the first ionisation limit \citep[also extracted from the NIST ASD;][]{NIST2023}. Both M1 and E2 transitions require parity to be conserved. M1 transitions can only possess variations in quantum $J$ number, $\Delta J = 0, \pm 1$ (but not $J = 0 \leftrightarrow 0$), while E2 transitions can only have $\Delta J = 0, \pm 1, \pm 2$ (but not $J = 0 \leftrightarrow 0$, $0 \leftrightarrow 1$ or $\sfrac{1}{2} \leftrightarrow \sfrac{1}{2}$). We apply these additional restrictions to our line lists, and discard any transitions that do not obey these criteria. Finally, for the transitions that possess level information expressed in the $LS$-coupling formalism, we apply the $LS$-coupling rules, and discard transitions that do not comply.

In a nebular regime, the emitted line luminosity ($L_{\rm em}$) for a transition can be expressed by:
\begin{equation}
    L_{\rm em} = A_{\textsc{ul}} \cdot N_{\textsc{u}} \left( \frac{h c}{\lambda_{\rm vac}} \right),
    \label{EQN: Emitted line luminosity}
\end{equation}
where $A_{\textsc{ul}}$ and $\lambda_{\rm vac}$ are the Einstein $A$-value and vacuum wavelength of the transition, respectively, and $N_{\textsc{u}}$ is the number of atoms/ions excited to the upper level. As our synthetic line lists do not possess any estimates for intrinsic line strengths, we assume equal Einstein $A$-coefficients for all transitions ($A_{\textsc{ul}} = 1$\,s$^{-1}$),\footnote{This represents the characteristic value for forbidden line strength invoked in this work; in reality, there will be a spread in \Aval\ for the lines shortlisted here, which can only be constrained by future atomic data generation.} meaning the \textit{relative} line luminosities for a species depend only on $N_{\textsc{u}}$ and $\lambda_{\rm vac}$.

We estimate level populations using the Boltzmann equation:
\begin{equation}
    N_{\textsc{u}} = N_{\textsc{t}} \left( \frac{g_{\textsc{u}}}{Z} \right) \cdot {\rm exp}\left[ {-\frac{E_{\textsc{u}}}{k_{\textsc{b}} T}} \right],
    \label{EQN: Boltzmann excitation}
\end{equation}
where $N_{\textsc{t}}$ is the total number of atoms or ions, $g_{\textsc{u}}$ is the statistical weight of the upper level, $Z$ is the LTE partition function, $E_{\textsc{u}}$ is the energy of the upper level, and $T$ is the temperature. However, we cannot directly constrain $N_\textsc{u}$ without estimates for the ejecta mass, composition and ionisation, and so we instead compute relative level populations, which only depend on $g_\textsc{u}$, $E_\textsc{u}$ and $T$. We compute our level population estimates for three representative temperature cases:
\begin{itemize}
    \item $T = 1000$\,K, motivated by the continuum $T$ estimates for the \jwst\ NIRSpec spectra of \vfi\ (see Section~\ref{SEC: Spectral fitting - AT2023vfi}),
    \item $T = 3000$\,K, motivated by continuum $T$ estimates for the late-time \xsh\ spectra of \gfo\ \citep[see \eg,][]{Waxman2018},
    \item $T = 5000$\,K, which represents the proposed typical temperature for nebular phase KN ejecta \citep[see \eg,][]{Hotokezaka2021}.
\end{itemize}

With these relative level populations, we compute luminosity estimates for each transition (as in Equation~\ref{EQN: Emitted line luminosity}), normalised to the line with the largest luminosity that has a wavelength within $\lambda = 1 - 5$\,\micron\ (the approximate range of rest wavelengths of \vfi\ sampled by the \jwst\ NIRSpec observations).\footnote{Note the slightly different approach to that presented by \cite{Gillanders2023_PaperII}. There, they presented the relative upper level population as a representation for transition strength.} Since we are interested in the forbidden transitions we expect to be most prominent for each species, we cut any from our shortlists that have a line strength $< 0.01$ that of the strongest line in our wavelength range.

Finally, we have estimates for the centroids of the observed emission features of \vfi\ (and \gfo; see Section~\ref{SEC: Spectral fitting}). We report a candidate transition for these features if it satisfies the following criteria:
\begin{itemize}
    \item It is expected to be (one of) the strongest emission transition(s) for the species,
    \item It does not possess a large number of other transitions that are also expected to be prominent, but which do not correspond to any observed emission,
    \item It lies within a 0.04\,c Doppler width of the measured centroid of the line (justified in Section~\ref{SEC: Line ID search - Wavelength windows for line search}),
    \item The upper energy level of the transition has $E_{\textsc{u}} \leq 4$\,eV.\footnote{We set this deliberately high threshold to account for any non-thermal and non-LTE excitation effects.}
\end{itemize}
While this approach does come with some uncertainty, it is capable of returning a subset of viable candidate line identifications \citep[as previously demonstrated for the case of \gfo; see][]{Gillanders2023_PaperII}, which we present in Section~\ref{SEC: Line ID search - Candidate ions}. Note that for this analysis we treat each ion independently; \ie, we have made no assumptions regarding the ionisation of the ejecta material.

We also note that our approach and results presented here are easily updated when Einstein $A$-values become available. Due to how we have performed our analysis, one can simply multiply our relative line strength estimates by the $A$-values to get updated line strengths (for LTE level populations at $T = 3000$\,K).

\subsection{Wavelength windows for line search} \label{SEC: Line ID search - Wavelength windows for line search}

In \cite{Gillanders2023_PaperII}, a transition was considered to be coincident with the observed emission features if it lies within a Doppler width of 0.02\,c from the measured centroid of the emission features. Here we relax this tolerable window to 0.04\,c (see Section~\ref{SEC: Line ID search - Methodology}) based on a number of points, presented below.

Our approach assumes that the observed emission features are dominated by emission from a single transition (or a small subset of transitions) from a single ion, centred on (or close to) the rest wavelengths of the lines. As such, our shortlisting approach suffers from two limitations.

First, we are not able to fully capture the effects of features being offset from their rest wavelengths. In nebular phase spectra of SNe, it is common to see emission features centred on their rest wavelengths, but we note there are a number of exceptions to this. In some fast-fading transients, significant offsets have been observed for the [Ca\,\II] doublet transitions. \cite{Foley2015} study the properties of 13 Ca-strong transients, and estimate the velocity shifts present in the nebular-phase [Ca\,\II] spectral features. These values vary across the sample, but are on the order of $\sim 1000$\,\kms\ (0.003\,c). The largest measured shift is $-1730$\,\kms\ ($-0.006$\,c), and belongs to \SNxx{2012hn}.

Other transients aside from Ca-strong SNe also exhibit velocity shifts. \ATxx{2018kzr} was a rapidly declining transient suggested to be produced by the merger of an oxygen--neon white dwarf star with either a neutron star or a (stellar mass) black hole (\citealt{McBrien2019, Gillanders2020}; see also \citealt{Bobrick2022}). The late-time spectra obtained for \ATxx{2018kzr} possessed prominent Ca\,\II\ NIR triplet features, from which a velocity shift of $-2300$\,\kms\ ($-0.008$\,c) was measured \citep{Gillanders2020}. We note that observing a shift in this WD--NS/BH merger system is relevant, as \thisGRB\ is also likely produced by a merger system.

Finally, \SNxx{2019bkc} is an unusual Ca-strong transient with an extremely rapid evolution \citep{Chen2020, Prentice2020}. Notably, the late-time spectral observations presented by \cite{Prentice2020} exhibit extremely blueshifted Ca\,\II\ and [Ca\,\II] features. \cite{Prentice2020} measure these offsets and record blueshifts for the Ca\,\II\ NIR triplet and [Ca\,\II] doublet transitions of \mbox{$10000 - 12000$\,\kms} ($0.033 - 0.040$\,c). \cite{Prentice2020} note that the late-time spectra also plausibly contain two O\,\I\ emission features ($\lambda \lambda 7772$, 9263) which exhibit a $\sim 10000$\,\kms\ (0.033\,c) blueshift.

With the exception of the tentative O\,\I\ features in \SNxx{2019bkc}, it seems that Ca\,\II\ is the ion that predominantly exhibits velocity offsets in explosive transient observations. This may be due to the high opacity associated with the NIR triplet, and if so, indicates that these observed offsets result from some type of opacity effect, rather than bulk velocity or ejecta inhomogeneity effects.

Clearly it is possible, at least in some cases, for feature offsets to be observed in late-time spectra of explosive transients. The magnitude of any feature offsets in late-time KN spectra is hard to quantify given the lack of observed systems and the lack of robust feature identifications, with one notable exception. \cite{Gillanders2023_PaperII} present analysis of the post-photospheric phase spectra of \gfo, with a focus on fitting the prominent emission features at late times. There they measure the centroid of a prominent emission feature at $\sim 1.08$\,\micron\ in the +7.4\,d \xsh\ spectrum of \gfo. Assuming this feature is dominated by late-time Sr\,\II\ NIR triplet emission, they estimate that the feature possesses a redshifted offset velocity of 9800\,\kms\ (0.033\,c).

Given the set of example cases above, clearly it is possible for some systems to produce emission features somewhat deviant from the rest wavelengths of the transitions powering them. Such effects in the case of \vfi\ may impact our ability to infer the correct line identifications for the observed emission features, as in Section~\ref{SEC: Line ID search - Candidate ions}.

Second, our assumption that the emission features are dominated by emission from a single ion may impact our inferences of the true feature centroids. There are expected to be many forbidden transitions belonging to the heavy \rpro\ species that have been synthesised, and as such line blending may be common. Quantitative modelling of KN spectra has shown that the assumption of a single strong emitter being completely responsible for emission features may be overly simplistic \citep[see \eg,][]{Kasen2017, Pognan2023, Shingles2023}. However, without the ability to undertake detailed modelling of KNe at late times, with full atomic line lists for all relevant \rpro\ elements, we cannot quantify the likelihood of this scenario in the case of \vfi.

To account for the above effects, we set our tolerable window to a Doppler width of 0.04\,c \citep[compared to the 0.02\,c tolerable window of][]{Gillanders2023_PaperII} and perform a search for candidate transitions, which we present below (Section~\ref{SEC: Line ID search - Candidate ions}). This approach allows us to capture a larger subset of transitions that may contribute to these features (allowing for the effects above) without having to undertake a quantitative detailed investigation; such a study with detailed radiative transfer is earmarked for future investigation. In the figures showing candidate transitions (\eg, Figure~\ref{FIG:TeI+II+III}) we highlight both the 0.02 and 0.04\,c tolerable windows separately for clarity.

\subsection{Candidate ions} \label{SEC: Line ID search - Candidate ions}

The transitions we recover for our analysis that can plausibly power the observed emission features in the \jwst\ NIRSpec spectra of \vfi\ are summarised in Table~\ref{TAB:Candidate lines}. We present the estimated relative strength of all transitions for the $T = 3000$\,K case, and where variations in $T$ significantly impact our results, we discuss the potential implications. We recover a list of possible contributing species, which we discuss in turn below. Where we present the number of transitions recovered by our analysis, this corresponds to all transitions that satisfy all steps of our shortlisting process, as listed in Section~\ref{SEC: Line ID search - Methodology}.

\subsubsection{Candidate transitions for the 2.0218\,\micron\ observed feature} \label{SEC: Line ID search - Candidate ions - 2.0}

\begin{itemize}

    \SpeciesX{56}{Ba}{ii} We recover two transitions for [Ba\,\II], with wavelengths of 17622 and 20518\,\AA\ (see Figure~\ref{FIG:BaII+RaII}). There is no observational evidence for the stronger 17622\,\AA\ transition. At lower $T$, the 20518\,\AA\ line becomes the dominant transition, but we still expect the 17622\,\AA\ line to persist.

    \SpeciesX{}{Ho}{iv} Seven lines recovered from our shortlisting; the most prominent has a wavelength of 19806\,\AA, and lies within our 0.04\,c tolerable window (see Figure~\ref{FIG:HoIV}).

    \SpeciesX{}{Er}{i} We recover 46 lines, but many of these lie close to our lower bound on relative strength, and so are expected to be $\sim$negligibly weak. We identify a strong line at a wavelength of 19860\,\AA, just within our tolerable threshold for the 2.0218\,\micron\ emission feature (see Figure~\ref{FIG:ErI+II+III}). The next most prominent transition lies at 14371\,\AA, for which we do not see evidence in the observations.

    \SpeciesX{}{Er}{ii} We recover 51 lines in our analysis, but as in the Er\,\I\ case, many of these are expected to be $\sim$negligibly weak. There are three strong lines at 19483, 21312 and 20148\,\AA, which are coincident with the 2.0218\,\micron\ (19483, 20148\,\AA) and 2.1874\,\micron\ features (21312\,\AA). There are four other prominent lines between \mbox{$\sim 1.4 - 1.5$\,\micron}, for which we do not see prominent emission in the observations (see Figure~\ref{FIG:ErI+II+III}).

    \SpeciesX{}{Er}{iii} Four lines satisfy our cuts, with the strongest having a wavelength of 19678\,\AA, coincident with the 2.0218\,\micron\ feature (see Figure~\ref{FIG:ErI+II+III}). We find a weaker (but still prominent) contaminant line at 14348\,\AA\ that does not match the data.

    \SpeciesX{}{Os}{i} We recover 22 lines, of which only three are expected to be prominent; these have wavelengths of 19440, 24042 and 36490\,\AA\ (see Figure~\ref{FIG:OsI}). The strongest line at $T = 3000$\,K is the 19440\,\AA\ transition, the one coincident with the 2.0218\,\micron\ feature. These other lines do not match the observational data. At higher $T$, the contaminant lines become less prominent (see Figure~\ref{FIG:Candidate ions - Temp evolution - OsI} to visualise line strength estimates across the range of explored $T$ values).

    \SpeciesX{}{Ir}{ii} We shortlist 12 lines from our analysis, but the line at 20886\,\AA\ is expected to be much stronger than the rest (see Figure~\ref{FIG:IrII}).

    \SpeciesX{}{Ra}{ii} One line recovered in our analysis within our wavelength cut of $1 \leq \lambda \leq 5$\,\micron; it has a wavelength of 20587\,\AA\ (see Figure~\ref{FIG:BaII+RaII}).

\end{itemize}

\subsubsection{Candidate transitions for the 2.1874\,\micron\ observed feature} \label{SEC: Line ID search - Candidate ions - 2.2}

\begin{itemize}

    \SpeciesX{36}{Kr}{iii} We shortlist two transitions, with wavelengths of 21986 and 18822\,\AA. The 21986\,\AA\ transition is predicted to be significantly more prominent than the 18822\,\AA\ transition, as needed to match observation (see Figure~\ref{FIG:KrIII}).

    \SpeciesX{}{Pd}{iii} We recover four transitions, of which two are expected to be most prominent; these have wavelengths of 21338 and 30967\,\AA\ (see Figure~\ref{FIG:PdIII}). The 21338\,\AA\ transition increases in relative strength with increasing $T$, but we find that across our range of $T$ values considered, the contaminant 30967\,\AA\ line is always dominant (see Figure~\ref{FIG:Candidate ions - Temp evolution - PdIII} to visualise line strength estimates across the range of explored $T$ values).

    \SpeciesX{47}{Ag}{iii} We recover a single transition, with a wavelength of 21696\,\AA, coincident with the 2.1874\,\micron\ feature (see Figure~\ref{FIG:AgIII}).

    \SpeciesX{}{Te}{i} Two transitions shortlisted from our analysis; these have wavelengths of 21049 and 21247\,\AA. The 21049\,\AA\ line lies exactly between the estimated centroids of the 2.0218 and 2.1874\,\micron\ features. The 21247\,\AA\ transition is coincident with the observed emission feature at 2.1874\,\micron\ in both the +29 and +61\,d spectra (see Figure~\ref{FIG:TeI+II+III}). The 21049\,\AA\ transition is $\sim 3 \times$ stronger, due to it having a $3 \times$ larger statistical weight for its upper level ($g_{\textsc{u}} = 3$ versus~1; the upper level energies are near-identical). Both transitions are too blue to satisfy our 0.02\,c tolerable window for the 2.1874\,\micron\ feature, but lie within our expanded 0.04\,c range.
    
    \SpeciesX{}{Te}{iii} Only three transitions are recovered from our analysis, with the strongest having a wavelength of 21050\,\AA\ (at an almost identical wavelength to the [Te\,{\sc i}] transition discussed above). This transition lies exactly between the estimated centroids of the 2.0218 and 2.1874\,\micron\ emission features. It makes our selection cut as it lies within the 0.04\,c range of the 2.1874\,\micron\ feature (see Figure~\ref{FIG:TeI+II+III}). The other two transitions have wavelengths of 12248 and 29290\,\AA, and there is no observational evidence for either of them in the observed data. While subdominant to the 21050\,\AA\ line, we still expect them to be detectable based on our analysis here.
    
    \SpeciesX{}{Er}{ii} See Section~\ref{SEC: Line ID search - Candidate ions - 2.0}. The 21312\,\AA\ line is coincident with the 2.1874\,\micron\ feature (see Figure~\ref{FIG:ErI+II+III}).
    
    \SpeciesX{}{Hf}{i} We recover 12 lines from our analysis, with the strongest three having wavelengths, $\lambda = 42433$, 21893 and 45229\,\AA, and \mbox{$L_{\rm em} = 1.0$}, 0.86 and 0.42 respectively (see Figure~\ref{FIG:HfI}). The 21893\,\AA\ line is coincident with the 2.1874\,\micron\ feature, while the 42433 and 45229\,\AA\ transitions lie blueward and slightly redward of the 4.4168\,\micron\ feature centroid (but lie within the 0.04\,c tolerable window), respectively.

    \SpeciesX{74}{W}{iii} We recover 13 transitions that satisfy our cuts, of which only five are expected to be prominent; these have wavelengths (ordered by line strength) of 22416, 44322, 24866, 45352 and 31003\,\AA\ (see Figure~\ref{FIG:WIII}). The strongest transition at $T = 3000$\,K (22416\,\AA) is coincident with the 2.1874\,\micron\ feature, and the 44322, 45352\,\AA\ lines are coincident with the 4.4168\,\micron\ feature. The other two prominent transitions do not match the observational data. 

    \SpeciesX{78}{Pt}{ii} Eight transitions recovered, but two are expected to be much more prominent than the others. These have wavelengths and strengths, $\lambda_{\rm vac} = 11877$, 21883\,\AA, and $L_{\rm em} = 1.0$, 0.69, respectively (see Figure~\ref{FIG:PtII}). The 11877\,\AA\ line is expected to be more prominent, although at higher $T$, the relative strength of the 21883\,\AA\ line increases.

\end{itemize}

\subsubsection{Candidate transitions for the 4.4168\,\micron\ observed feature} \label{SEC: Line ID search - Candidate ions - 4.4}

\begin{itemize}

    \SpeciesX{}{Se}{iii} We shortlist just two transitions, with wavelengths of 25401 and 45549\,\AA. There is no observational evidence in support of the 25401\,\AA\ line, but the 45549\,\AA\ line lies within the tolerable range of the 4.4168\,\micron\ feature (see Figure~\ref{FIG:SeIII}).

    \SpeciesX{}{Nb}{iv} Four lines are shortlisted, but the only one we expect to be prominent has a wavelength of 42637\,\AA\ (see Figure~\ref{FIG:NbIV}).
    
    \SpeciesX{}{In}{i} We recover a single line with a wavelength of 45196\,\AA\ (see Figure~\ref{FIG:InI}).
    
    \SpeciesX{}{Te}{ii} We recover two transitions, but only the 45466\,\AA\ line is expected to be prominent (see Figure~\ref{FIG:TeI+II+III}).
    
    \SpeciesX{58}{Ce}{iv} We recover only one transition with a wavelength of 44391\,\AA, coincident with the 4.4168\,\micron\ feature (see Figure~\ref{FIG:CeIV}).

    \SpeciesX{62}{Sm}{iv} We recover 26 lines in our analysis, with the strongest four having wavelengths of 35024, 37010, 39485 and 43656\,\AA\ (see Figure~\ref{FIG:SmIV}). The 43656\,\AA\ line is the most prominent transition across our range of $T$ values, but the contaminant lines increase in relative strength with increasing $T$ (see Figure~\ref{FIG:Candidate ions - Temp evolution - SmIV} to visualise the variation in relative line strength estimates across the range of explored $T$ values). We find we best match the data with low $T$, which reduces the prominence of the 35024, 37010 and 39485\,\AA\ lines, in addition to the grouping of contaminant lines between $\sim 1 - 2$\,\micron\ (see Figures~\ref{FIG:SmIV}~and~\ref{FIG:Candidate ions - Temp evolution - SmIV}). 

    \SpeciesX{}{Hf}{i} See Section~\ref{SEC: Line ID search - Candidate ions - 2.2} and Figure~\ref{FIG:HfI}. The 42433 and 45229\,\AA\ lines are coincident with the 4.4168\,\micron\ feature.

    \SpeciesX{74}{W}{iii} See Section~\ref{SEC: Line ID search - Candidate ions - 2.2} and Figure~\ref{FIG:WIII}. The second and fourth strongest transitions at $T = 3000$\,K (44322 and 45352\,\AA, respectively) are coincident with the 4.4168\,\micron\ feature. At lower $T$, the 44322\,\AA\ line becomes the most prominent transition (see Figure~\ref{FIG:Candidate ions - Temp evolution - WIII} to visualise line strength estimates across the range of explored $T$ values).

    \SpeciesX{89}{Ac}{i} Three transitions recovered, but only one is expected to be prominent; it has a wavelength of 44814\,\AA\ (see Figure~\ref{FIG:AcI}).

\end{itemize}

\subsection{Discussion} \label{SEC: Line ID search - Discussion}

From our line identification analysis, we have produced a shortlist of candidate species that may feasibly be contributing to the observed emission features in the +29 and +61~day \jwst\ NIRSpec spectra of \vfi. This list of ions includes the `light' ($30 \leq Z \leq 56$) \rpro\ species Se\,\III, Kr\,\III, Nb\,\IV, Pd\,\III, Ag\,\III, In\,\I, Te\,\I--\III\ and Ba\,\II, the lanthanide ($57 \leq Z \leq 71$) species Ce\,\IV, Sm\,\IV, Ho\,\IV\ and Er\,\I--\III, and the `heavy' ($72 \leq Z \leq 92$) \rpro\ species Hf\,\I, W\,\III, Os\,\I, Ir\,\II, Pt\,\II, Ra\,\II\ and Ac\,\I.

The majority of the lines that we shortlist here and expect to be most prominent correspond to fine-structure transitions between the ground terms of the species under investigation. These transitions are expected to be among the most prominent forbidden emission transitions, and are expected to dominate nebular phase emission in kilonovae \citep[see][]{Hotokezaka2022}.

While our shortlist of candidates seems long, we note that it provides a significant reduction to the list of possible ions that warrant prioritised detailed atomic data studies for advancing our understanding of KNe, specifically \vfi. Given the number of elements considered in this line identification search ($30 \leq Z \leq 92$) and four ionisation stages for each element, the initial list of ions under consideration was 252; our shortlist here contains just 23 ions. This analysis here represents an order-of-magnitude reduction in the ions that should be investigated first to more completely model the spectra of \vfi. As such, we propose the list above represents the ions that should be investigated to more fully understand the late-time emission present in \vfi.

While all of the above proposed species warrant further study, below we discuss a few ions of particular interest.

\subsubsectionRoman{[Te\,\I], [Te\,\II] {\it and} [Te\,\III]} \label{SEC: Line ID search - Discussion - Te}

\begin{figure}
    \centering
    \subfigure{\includegraphics[width=\linewidth]{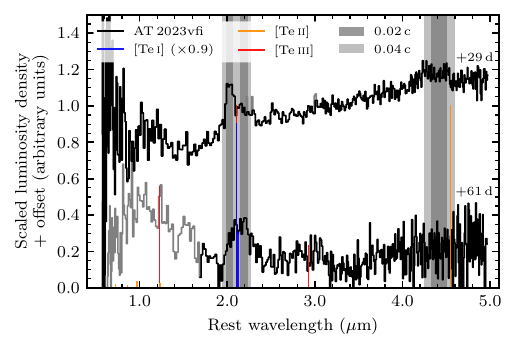}}
    \subfigure{\includegraphics[width=\linewidth]{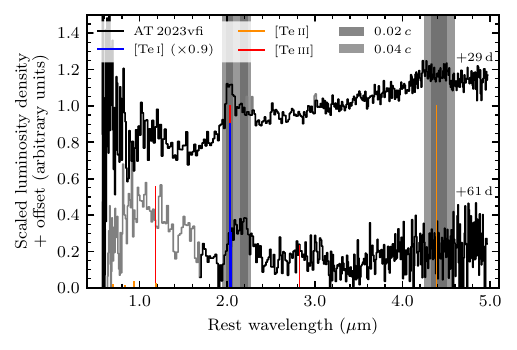}}
    \caption{
        \textit{Upper:} Comparison of our [Te\,\I], [Te\,\II] and [Te\,\III] emission line spectra and the (arbitrarily re-scaled and offset) +29 and +61\,d \jwst\ spectra of \vfi\ (annotated with their phase relative to the GRB trigger). The vertical bands represent the 0.02 and 0.04\,c tolerable windows considered for each feature (dark and light grey, respectively), where lines that lie within these bands are considered to plausibly match the observed emission features. Emission lines are plotted with their relative strengths (computed for \mbox{$T = 3000$\,K}) reflected by their prominence. The lines have been normalised for each ion such that the strongest line between $1 - 5$\,\micron\ has a luminosity value of 1; \ie, all ions have been treated independently, meaning the relative ionisation fraction of Te has not been taken into account. In this instance, we re-scale [Te\,\I] by a constant factor to improve clarity between the overlapping [Te\,\I]~$\lambda 21049$ and [Te\,\III]~$\lambda 21050$ lines.
        \textit{Lower:} Same as the upper panel, but with a constant blueshift of 0.035\,c applied to all emission lines of [Te\,\I], [Te\,\II] and [Te\,\III].
    }
    \label{FIG:TeI+II+III}
\end{figure}

Much has been made of the possible detection of Te (specifically Te\,\III) in the spectra of both \gfo\ and \vfi\ \citep[][]{Hotokezaka2023,Gillanders2023arxiv_GRB230307A, Gillanders2023_PaperII, Levan2024}. The line that has sparked such conclusions is the result of a [Te\,\III] fine-structure transition with a rest wavelength, $\lambda_{\rm vac} = 21050$\,\AA. This line has been observed in NIR spectra of two planetary nebulae by \cite{Madonna2018}, who quote a measured line position of 2.1019\,\micron. This transition lies redward of the observed 2.0218\,\micron\ emission feature, and blueward of the observed 2.1874\,\micron\ emission feature. It is shortlisted here in this work as it lies within the 0.04\,c tolerable window of the 2.1874\,\micron\ observed emission feature (see Figure~\ref{FIG:TeI+II+III} and Table~\ref{TAB:Candidate lines}). In fact, this transition lies exactly at the average of the two observed peaks ($\lambda_{\textsc{av}} = 21046$\,\AA). For Te\,\III\ to be responsible for producing either (or both) of these observed features, we would need some opacity, line blanketing or bulk velocity offset effects to explain why the observed peaks are shifted from the rest wavelength of the transition (see Section~\ref{SEC: Line ID search - Interpreting the 2.1um emission as a single feature}). While we cannot rule out such effects, and thus cannot rule out Te\,\III\ being the source of (one of) the emission feature(s) observed in both \gfo\ and \vfi, we also cannot definitively state that Te\,\III\ is the source. Thus, caution should be applied to this (and any other) proposed line identification until more robust evidence is presented (\eg, multiple emission features well-matched by several transitions from the same ion; \cf, Ca\,\II\ H\&K and NIR triplet transitions in SN spectra).

Another point of consideration with regards [Te\,\III] as a source of the observed emission is that we find that there are three transitions that we expect to be prominent, with wavelengths, $\lambda_{\rm vac} = 12248$, 21050 and 29290\,\AA. All three of these are transitions between the fine-structure levels in the ground term (5s$^2$5p$^2$~$^3$P), all of which we expect to be significantly populated. As such, all three of these transitions are expected to be somewhat prominent in our analysis, with the 21050\,\AA\ line being the strongest (see Figure~\ref{FIG:TeI+II+III}). There is no observational evidence for the 12248\footnote{There may be a noisy flux excess present in the +29\,d \jwst\ spectrum of \vfi\ that is coincident with the location of this 12248\,\AA\ transition; see the discussion in Section~\ref{SEC: Line ID search - Discussion - PtII}.} and 29290\,\AA\ lines. \cite{Madonna2018} present \Aval\ for a number of [Te\,\III] transitions, including the three shortlisted here in this work. They quote \mbox{$A_{\textsc{ul}} = 0.012$}, 1.19 and 0.52\,s$^{-1}$ for the 12248, 21050 and 29290\,\AA\ lines, respectively. With this additional information, it is clear that the 12248\,\AA\ line will be much weaker than our estimate here, and thus not detectable. 

\begin{figure*}
    \centering
    \includegraphics[width=0.8\linewidth]{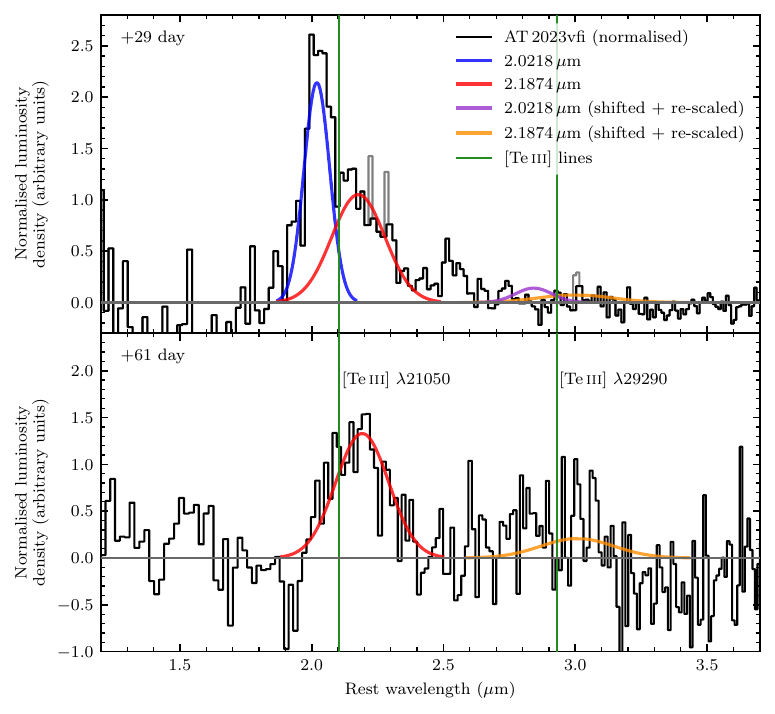}
    \caption{
        \textit{Upper:} +29~day \jwst\ spectrum of \vfi, continuum-subtracted (model AG+BB continuum; see Table~\ref{TAB:Continuum parameters}) and normalised (black), compared with our best-fitting model Gaussian fits for the 2.0218 and 2.1874\,\micron\ features (blue and red, respectively). The emission line profiles for the [Te\,\III] 29290\,\AA\ line are computed assuming these two emission features are powered by [Te\,\III] $\lambda 21050$ (see the main text), and are also plotted (purple and orange, respectively). The rest wavelengths of the [Te\,\III] $\lambda \lambda 21050$, 29290 lines are marked (vertical green lines). Here we invoke \Aval\ from \protect\cite{Madonna2018} to compute line luminosities (see the main text).
        \textit{Lower:} Same as the upper panel, but for the +61~day \jwst\ spectrum and the 2.1874\,\micron\ feature.
    }
    \label{FIG:TeIII 29290AA profiles}
\end{figure*}

\begin{figure*}
    \centering
    \includegraphics[width=0.8\linewidth]{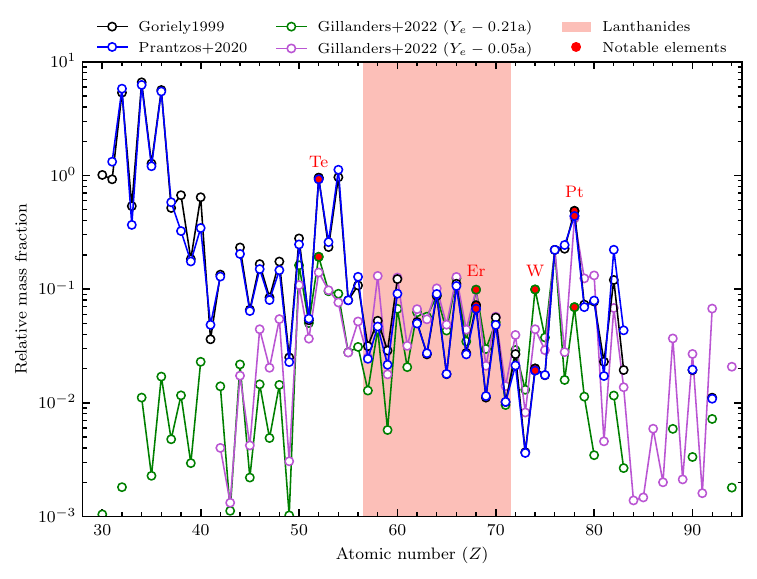}
    \caption{
        Mass fraction distributions of the \rpro\ elements in the Solar system, as estimated by \protect\cite{Goriely1999} and \protect\cite{Prantzos2020} (black and blue, respectively). We also show two of the composition profiles presented by \protect\cite{Gillanders2022_PaperI}, to illustrate the possible deviation from the Solar \rpro\ distribution for two characteristic \Ye\ regimes ($Y_e = 0.21, 0.05$; green and purple, respectively). All abundance patterns have been scaled such that they have identical relative mass fractions for Os ($Z = 76$). The lanthanide elements are highlighted by a vertical pink band, and we mark the locations of a number of notable elements (with red dots) discussed in the main text (Section~\ref{SEC: Line ID search - Discussion}).
    }
    \label{FIG:r-process abundances}
\end{figure*}

With our relative line strength estimates for [Te\,\III] and the measured luminosity estimates for the +29\ and +61\,d 2.0218 and 2.1874\,\micron\ emission features (see Table~\ref{TAB:Feature parameters}), we compute the expected emission profiles for the [Te\,\III] 29290\,\AA\ line. By assuming that the emission feature at 2.0218\,\micron\ is powered entirely by [Te\,\III] $\lambda 21050$, we can compute the emitted luminosity of the 29290\,\AA\ line from our relative strength estimates and the \Aval\ from \cite{Madonna2018}. We maintain the same $v_{\textsc{fwhm}}$ value, and assume the same offset as observed between the 21050\,\AA\ line and the 2.0218\,\micron\ emission feature ($\Delta \lambda = -832$\,\AA; \ie, the emission feature computed for 29290\,\AA\ is centred at 28458\,\AA). We also perform the same process for the 2.1874\,\micron\ emission feature at both +29 and +61\,d, and centre it on 30114\,\AA. These computed emission line profiles are presented in Figure~\ref{FIG:TeIII 29290AA profiles}.

We find that at $T = 5000$\,K, the expected emission from the 29290\,\AA\ line barely rises above the continuum noise at both +29 and +61~days; therefore, it seems plausible that our observations contain prominent [Te\,\III] emission from the 21050\,\AA\ line, but no detectable signal from the 12248 and 29290\,\AA\ lines. The relative strength of the 29290\,\AA\ line decreases with decreasing $T$, and so for the other temperatures explored in this work ($T = 1000$, 3000\,K), the 29290\,\AA\ line is expected to be even weaker than the profiles shown in Figure~\ref{FIG:TeIII 29290AA profiles}. \cite{Hotokezaka2023} find that this 29290\,\AA\ line is not prominent in their models for the $+7.4 - 10.4$\,d \xsh\ spectra of \gfo\ (assuming $T_e = 2000$\,K), although they note the line becomes more prominent with increasing $T$. Their models do have a prominent feature close to this wavelength, but it is attributed to more prominent emission from [Os\,\II]. We agree with the conclusion of \cite{Hotokezaka2023} that higher $T$ is necessary for this line to become prominent. Here we also conclude that it is likely not detectable above the noise in the existing \jwst\ spectra of \vfi. However, we estimate that if the upper level population was $\sim 3 \times$ higher than that computed assuming LTE level populations at \mbox{$T = 5000$\,K}, the 29290\,\AA\ line would become detectable.

For the [Te\,\I] case, we find that there are two transitions that are expected to be prominent, both of which lie within the 0.04\,c tolerable window of the $2.1874$\,\micron\ feature; these have $\lambda_{\rm vac} = 21049$ and 21247\,\AA\ (see Figure~\ref{FIG:TeI+II+III} and Table~\ref{TAB:Candidate lines}). As in the [Te\,\III] case, these lines do not exactly match the feature centroids at either $2.0218$ or $2.1874$\,\micron. We note that the 21049\,\AA\ line exactly matches the average of the two observed emission peaks. The same arguments for [Te\,\III] apply here too -- perhaps some opacity, bulk offset or line blending effect is conspiring to shift the apparent feature peak away from the rest wavelengths of these lines (see Section~\ref{SEC: Line ID search - Interpreting the 2.1um emission as a single feature}). Also, we note that we typically would not expect significant neutral Te to exist alongside doubly ionised material in an LTE regime; however, non-LTE and non-thermal effects can cause multiple ion stages to co-exist, and so it remains possible that some combination of Te\,\I, \II\ and \III\ all exist in some significant fraction and contribute to the observed emission features \citep[see][]{Pognan2022}. \cite{Hotokezaka2023} suggest that the contribution of [Te\,\I] emission compared to [Te\,\III] in \gfo\ should be sub-dominant. 

Finally, Te\,\II\ only possesses a single strong transition in our analysis, with $\lambda_{\rm vac} = 45466$\,\AA. We find this line lies within the 0.04\,c tolerable window of the emission feature at 4.4168\,\micron. It is too red to exactly match the emission feature, but as noted above, this offset may be explainable. Clearly, if Te\,\II\ exists in any significant quantity in the ejecta, the most promising method of detection would be emission centred near 45466\,\AA. Also, it may be possible in some regimes for singly ionised Te to co-exist alongside doubly ionised and/or neutral Te. In this case, the identification of emission features belonging to multiple ionisation stages of Te is notable, as, if true, it would mark the first instance of multiple features in KN spectra being linked to the same element.

We note that all proposed line identifications for Te\,\I, \II\ and \III\ are systematically too red to exactly match the centroids of the 2.0218 and 4.4168\,\micron\ features. However, a common blueshift of \mbox{$\approx 0.025 - 0.045$\,c} shifts all prominent Te lines to match (\ie, to lie within the 0.02\,c window) the 2.0218 and 4.4168\,\micron\ feature centroids (see Figure~\ref{FIG:TeI+II+III}). This consistent offset indicates the presence of some residual opacity effect that impacts Te across the $\sim 2.0 - 4.5$\,\micron\ wavelength range, or some bulk velocity/inhomogeneity in the ejecta material that results in Te disproportionally moving towards the observer (see Section~\ref{SEC: Line ID search - Interpreting the 2.1um emission as a single feature}).

To check whether the identification of [Te\,\III] $\lambda 21050$ is reasonable, we estimate the required mass of Te\,\III\ needed to produce the measured luminosity of the observed emission features (reported in Table~\ref{TAB:Feature parameters}). Assuming the 2.0218\,\micron\ (2.1874\,\micron) feature at +29\,d is dominated by emission from the [Te\,\III] $\lambda 21050$ line, we use Equations~\ref{EQN: Emitted line luminosity}~and~\ref{EQN: Boltzmann excitation} to compute the approximate mass of Te\,\III\ ($M_{\textrm{Te\,\textsc{iii}}}$) needed to power the feature to be \mbox{$M_{\textrm{Te\,\textsc{iii}}} \approx 3.3 \times 10^{-5}$\,\msun} \mbox{($M_{\textrm{Te\,\textsc{iii}}} \approx 3.9 \times 10^{-5}$\,\msun)}. The mass needed to power the 2.1874\,\micron\ emission at +61\,d is \mbox{$M_{\textrm{Te\,\textsc{iii}}} \approx 1.3 \times 10^{-5}$\,\msun}. \cite{Levan2024} estimate the Te\,\III\ mass in the line-forming region of the ejecta of \vfi\ needed to power the observed flux at $\sim 2.1$\,\micron\ at +29\,d to be $\approx 10^{-3}$\,\msun. This value is somewhat higher (factor of $\approx 30$) than our estimate, for two main reasons. First, we have estimated the mass of Te\,\III\ needed to power one component of the blended $\sim 2.1$\,\micron\ emission, whereas \cite{Levan2024} estimate the Te\,\III\ mass by considering the entire flux density across the wavelength range of the blended emission feature ($2.02 - 2.48$\,\micron, observer frame). Therefore, they are estimating a larger mass of Te\,\III\ since they are fitting a larger line luminosity ($L_{\rm line} \approx 3 \times 10^{38}$\,\ergs, versus our estimate of $6.4 \times 10^{37}$\,\ergs; see Table~\ref{TAB:Feature parameters}). Second, we have assumed LTE level populations, whereas \cite{Levan2024} have considered non-LTE effects for their calculation. Specifically, \cite{Levan2024} utilise collisional information to estimate level populations considering both collisional and radiative effects, and they invoke an ionisation balance for Te to match that presented by \cite{Pognan2022}. Our LTE level populations may overestimate the upper level population, leading to a smaller inferred mass than that estimated by \cite{Levan2024}.

Although visually the observed features and the rest wavelengths of the Te\,\I--\III\ transitions do not perfectly match, there is clear motivation for a more detailed investigation into the possible identification of Te. Te is an ideal candidate element for further study, since it possesses multiple transitions from multiple ion stages that match the observed emission features, and it is expected to be abundant across a range of \rpro\ scenarios \citep[owing to it being an even-$Z$ element that lies at the top of the second \rpro\ peak; see Figure~\ref{FIG:r-process abundances} and \eg,][]{Seeger1965, Goriely1999, Sneden2008, Thielemann2011, Thielemann2017, Prantzos2020, Cowan2021}.

In Figure~\ref{FIG:r-process abundances} we present the Solar \rpro\ abundance profiles, as estimated by \cite{Goriely1999} and \cite{Prantzos2020}. In both, Te is the seventh most abundant element (and is the second most abundant element for $Z \geq 38$). We also present two composition profiles from \cite{Gillanders2022_PaperI}, which show the predicted composition for binary neutron star (BNS) merger ejecta with characteristic \Ye\ values of 0.21 and 0.05 \citep[derived from a nucleosynthetic calculation based on the dynamical mass ejection of a realistic hydrodynamical simulation of a BNS merger;][]{Goriely2011, Goriely2013, Goriely2015, Bauswein2013}. For the \YeTwoOne\ and \YeZeroFive\ models, Te is the second and third most abundant element (across all $Z$), respectively. Clearly, Te is an intrinsically abundant \rpro\ element, and so it seems natural to expect an element as abundant as Te to be detectable in the observed spectra.

To summarise, the [Te\,\III] 21050\,\AA\ line lies exactly between the 2.0218 and 2.1874\,\micron\ emission features (see Figure~\ref{FIG:TeI+II+III}), and we expect this line to be prominent. Thus, we agree with \cite{Levan2024} and \cite{Gillanders2023arxiv_GRB230307A} that the [Te\,\III] 21050\,\AA\ transition is the current most likely candidate for producing (one of) the emission feature(s) at $\sim 2.1$\,\micron\ in \vfi. However, given the blended line profile at +29\,d and the evolution to +61\,d (with an apparent different profile), and the non-negligible offset from both the 2.0218 and 2.1874\,\micron\ components, we stop short of confirming the presence of [Te\,\III] emission in the \jwst\ spectroscopic observations of \vfi. Further work is needed before drawing such a conclusion. Specifically, focus should be placed on obtaining detailed atomic data for the lowest few ion stages, including estimates for transition strengths and collisional information. These data could then be used for quantitative non-LTE spectral modelling of KNe \citep[as in, \eg,][]{Pognan2023}, to confirm whether the observed offsets and double profile can be explained with opacity effects, or bulk velocity flows (see Section~\ref{SEC: Line ID search - Interpreting the 2.1um emission as a single feature}). This would also confirm whether the other transitions of Te\,\III\ (and of Te\,\I~and~Te\,\II) are also expected to be prominent and to contribute to the observed spectra.

\subsubsectionRoman{[Ba\,\II] {\it and} [Ra\,\II]} \label{SEC: Line ID search - Discussion - BaII + RaII}

\begin{figure}
    \centering
    \includegraphics[width=\linewidth]{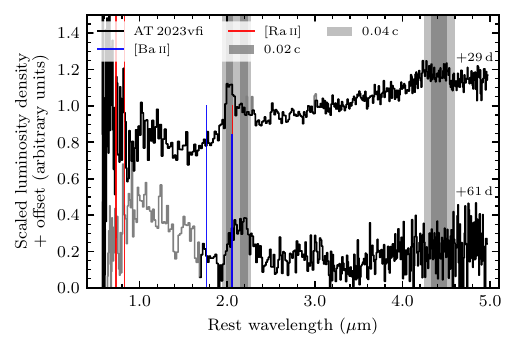}
    \caption{
        Same as Figure~\ref{FIG:TeI+II+III}, but here we present the emission line spectra for [Ba\,\II] and [Ra\,\II].
    }
    \label{FIG:BaII+RaII}
\end{figure}

\begin{figure}
    \centering
    \includegraphics[width=\linewidth]{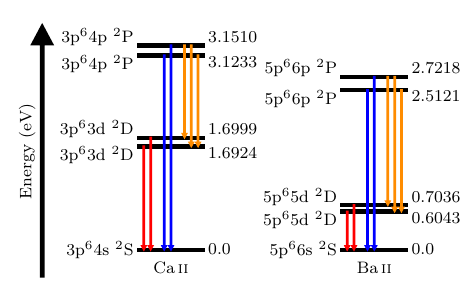}
    \caption{
        Energy level diagrams for the first five levels of Ca\,\II\ and Ba\,\II. The levels are annotated with their configuration and term (left), and their corresponding energies (in eV; right). Close-lying levels have been slightly offset for clarity. For the Ca\,\II\ ion, different sets of transitions have been marked with coloured arrows. The [Ca\,\II] doublet, the H\&K resonance doublet and the NIR triplet transitions are marked with red, blue and orange arrows, respectively. The equivalent transitions in the Ba\,\II\ ion are marked in the same manner. All lines are presented as emission transitions.
    }
    \label{FIG:BaII energy levels}
\end{figure}

We find that [Ba\,\II] possesses two prominent forbidden transitions, with $\lambda_{\rm vac} = 17622$ and 20518\,\AA\ (see Figure~\ref{FIG:BaII+RaII}). \cite{Levan2024} also shortlist [Ba\,\II] 20518\,\AA\ as a possible source of the 2.0218\,\micron\ feature in \vfi. These two transitions originate from the 5p$^6$5d~$^{2}$D levels, and de-excite to the ground state (5p$^6$6s~$^{2}$S). These doublet transitions, and the Ba\,\II\ system as a whole, is analogous to the well-studied Ca\,\II\ system (as illustrated in Figure~\ref{FIG:BaII energy levels}), as they are both Group~2 elements. In the Ca\,\II\ case, the forbidden doublet transitions (with $\lambda_{\rm vac} = 7293.5$ and 7325.9\,\AA) are routinely detected as a blend of prominent emission in late-phase supernova spectra \citep[see \eg,][]{LiMcCray93,Jerkstrand2012,Kumar2022}. The close-lying upper levels in the Ca\,\II\ case results in small wavelength spacing between the doublet transitions ($\Delta \lambda \simeq 30$\,\AA; see Figure~\ref{FIG:BaII energy levels}), and so they appear as a single, blended emission feature. In the Ba\,\II\ case however, the larger upper level separation results in well-spaced doublet lines, separated by $\Delta \lambda \simeq 2900$\,\AA\ (see Figure~\ref{FIG:BaII energy levels}).

We find that both of these [Ba\,\II] lines may be prominent and observable as individual strong emission features, centred at 17622 and 20518\,\AA\ (see Figure~\ref{FIG:BaII+RaII}). By considering the statistical weights and upper level populations, and estimating the resultant line luminosities, we find that the 17622\,\AA\ transition should be more luminous than its counterpart at 20518\,\AA\ by a factor of $\approx 1.2$ (for $T = 3000$\,K). Hence if one of these lines is present in the observations, then the other must also be present and observable (barring extreme line blanketing effects). Thus, although the [Ba\,\II] 20518\,\AA\ transition has a wavelength coincident with the emission feature at 2.0218\,\micron, we ultimately rule Ba\,\II\ out as a candidate ion, since there is no evidence in the observational data for any emission at 17622\,\AA.

We find good agreement between the [Ra\,\II] 20587\,\AA\ line and the observed 2.0218\,\micron\ emission feature (see Figure~\ref{FIG:BaII+RaII}). However, Ra\,\II\ is also a Group~2 element, and thus a homologue of Ba\,\II\ and Ca\,\II. We find that the equivalent transitions to the [Ca\,\II] doublet (which have $\lambda_{\rm vac} = 7276.4, 8275.2$\,\AA\ in the Ra\,\II\ case) are expected to be $\sim 1000 \times$ stronger than this 20587\,\AA\ transition. Although the observational data are very noisy $\lesssim 1$\,\micron, it seems unlikely that such prominent emission from this [Ra\,\II] doublet would not be observationally detected in the \vfi\ spectra. Therefore, we ultimately rule out Ra\,\II\ as a viable candidate ion for the 2.0218\,\micron\ feature.

\subsubsectionRoman{[Er\,\I], [Er\,\II] {\it and} [Er\,\III]} \label{SEC: Line ID search - Discussion - Er}

\begin{figure}
    \centering
    \includegraphics[width=\linewidth]{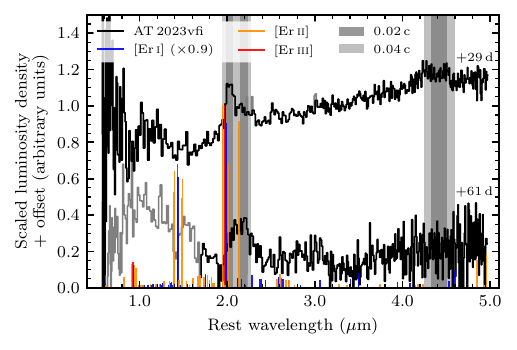}
    \caption{
        Same as Figure~\ref{FIG:TeI+II+III}, but here we present the emission line spectra for [Er\,\I], [Er\,\II] and [Er\,\III]. We arbitrarily re-scale [Er\,\I] by a constant factor to improve clarity with overlapping lines.
    }
    \label{FIG:ErI+II+III}
\end{figure}

We find that Er\,\I, \II\ and \III\ all have their strongest lines coincident with the 2.0218 and 2.1874\,\micron\ emission features (see Figure~\ref{FIG:ErI+II+III} and Table~\ref{TAB:Candidate lines}). The [Er\,\I] $\lambda 19860$ and [Er\,\II] $\lambda 20148$ lines lie within the 0.02\,c window for the 2.0218\,\micron\ feature, while all other matching lines ([Er\,\II]~$\lambda \lambda 19483, 21312$, [Er\,\III] $\lambda 19678$) all lie at the blue edges of the 0.04\,c tolerable windows for the 2.0218 and 2.1874\,\micron\ features. This grouping of lines, belonging to the lowest three ion stages of Er, all plausibly contribute to the emission features at 2.0218 and 2.1874\,\micron.

However, we find a number of contaminant lines clustered at \mbox{$\sim 1.4 - 1.5$\,\micron} (see Figure~\ref{FIG:ErI+II+III}), which are expected to also be prominent. There is no evidence for any excess emission at this wavelength range in either the +29 or +61\,d spectra. Although this may be explained by some line blanketing effect from other species, which would act to suppress the emission from these lines, it is not clear if such extreme blanketing is possible at the wavelengths and phases under consideration. It is also possible that our estimated \Aval\ and/or upper level populations for these contaminant transitions are over-estimates, such that these lines are weaker than we have calculated. As such, we conclude that quantitative modelling (with more complete atomic data) is needed to explore the possible effects of line blanketing, and whether some combination of Er\,\I--\III\ are the species responsible for the 2.0218 and 2.1874\,\micron\ emission features in the \jwst\ spectra of \vfi. Despite this apparent discrepancy with the observational data, we find that Er is the lanthanide most likely to have multiple ions contributing to the emission features in \vfi. Thus, we highlight Er\,\I--\III\ as priority lanthanide species that warrant further detailed atomic data study.

\subsubsectionRoman{[W\,\III]} \label{SEC: Line ID search - Discussion - WIII}

\begin{figure}
    \centering
    \subfigure{\includegraphics[width=\linewidth]{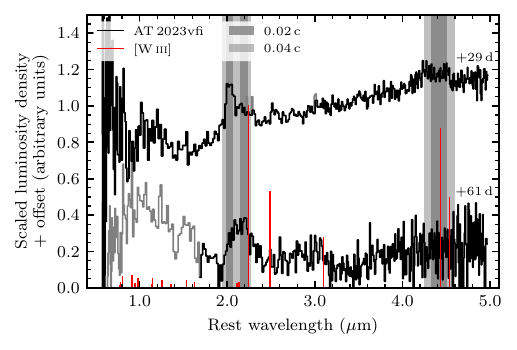}}
    \subfigure{\includegraphics[width=\linewidth]{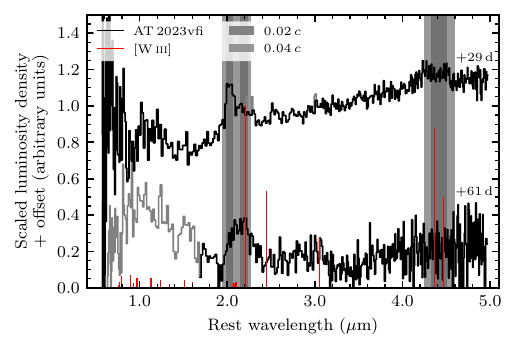}}
    \caption{
        \textit{Upper:} Same as Figure~\ref{FIG:TeI+II+III}, but here we present the emission line spectrum for [W\,\III].
        \textit{Bottom:} Same as the upper panel, but with a constant blueshift of 0.015\,c applied to all emission lines of [W\,\III].
    }
    \label{FIG:WIII}
\end{figure}

We find that W\,\III\ has a number of prominent lines in close agreement with the observed emission features of \vfi; these include the 22416, 44322 and 45352\,\AA\ transitions (see Figure~\ref{FIG:WIII} and Table~\ref{TAB:Candidate lines}). The 44322\,\AA\ line lies within the 0.02\,c window for the 4.4168\,\micron\ feature. This line is the most prominent in our analysis at $T = 1000$\,K, and second most prominent at $T = 3000$\,K (see Figure~\ref{FIG:Candidate ions - Temp evolution - WIII}). The most prominent line at $T = 3000$\,K is the 22416\,\AA\ transition, which lies within the 0.04\,c range of the 2.1874\,\micron\ transition. The third line (45352\,\AA) lies within the 0.04\,c window for the 4.4168\,\micron\ feature. There are two other prominent lines (at $T = 3000$\,K) that do not match any observed emission features (with wavelengths of 24866 and 31003\,\AA).

If \vfi\ underwent heavy \rpro\ nucleosynthesis, and material was synthesised up to the third \rpro\ peak, then we may expect a significant quantity of W to be produced. Although W ($Z = 74$) is not that abundant relative to the third \rpro\ peak elements Os, Ir, Pt and Au ($Z = 76$, 77, 78 and 79, respectively) in the Solar \rpro\ distribution \citep[][see Figure~\ref{FIG:r-process abundances}]{Goriely1999, Prantzos2020}, individual systems undergoing \rpro\ nucleosynthesis can deviate substantially from this average \rpro\ distribution. This is evidenced by the composition profiles presented by \eg, \cite{Gillanders2022_PaperI}, where W constitutes a few~per~cent of the mass fractions, at least for low \Ye\ regimes. The \YeTwoOne\ and \YeZeroFive\ composition profiles of \cite{Gillanders2022_PaperI} are presented in Figure~\ref{FIG:r-process abundances}, and there we see that W is relatively more abundant than in the Solar \rpro\ distributions. In particular, for the \YeTwoOne\ composition profile, W is predicted to be the fifth most abundant \rpro\ element across all $Z$.

If the bulk of W is doubly ionised, then the strongest traces of this element at late phases are likely to be the lines reported here. While the 22416 and 45352\,\AA\ lines require some explanation for the small offset between the observed 2.1874 and 4.4168\,\micron\ features and their rest wavelengths (see the discussion above for the Te case; \ie, Section~\ref{SEC: Line ID search - Discussion - Te}), the 44322\,\AA\ line agrees well with the observed peak of the 4.4168\,\micron\ feature. Perhaps this feature is powered by [W\,\III] $\lambda 44322$ emission and the other transitions contribute to their respective coincident emission components, but are not the dominant transition, hence the apparent offset. Alternatively, some velocity offset, as in the Te case (see Section~\ref{SEC: Line ID search - Discussion - Te}) could explain the observed feature offsets. Applying a blueshift of $\approx 0.01 - 0.02$\,c aligns the 22416, 44322 and 45352\,\AA\ transitions with the observed centroids of the 2.1874 and 4.4168\,\micron\ emission features (\ie, they lie within the 0.02\,c windows of the respective features; see Figure~\ref{FIG:WIII}). Detection of [W\,\III] emission has been previously reported in kilonova spectra; \cite{Hotokezaka2022} propose that the [W\,\III] $\lambda \lambda 44322$, 45352 transitions may be responsible for powering the late-time flux excess recorded by \spitzer\ for \gfo, while \cite{Levan2024} have previously proposed that [W\,\III] $\lambda 44322$ may contribute to the flux excess at $\sim 4.5$\,\micron\ in the +29\,d \jwst\ spectrum of \vfi.

The identification of three prominent transitions, all belonging to W\,\III, coincident with two of the observed emission components in \vfi\ is encouraging. This presents a potential identification whereby multiple features may be attributable to several transitions belonging to a single ion. Further investigation is needed for W\,\III; specifically, a complete list of forbidden transitions with computed intrinsic line strengths, as well as collisional rate estimates are needed to enable detailed non-LTE modelling (as noted above for the case of Te; see Section~\ref{SEC: Line ID search - Discussion - Te}).

\subsubsectionRoman{[Pt\,\II]} \label{SEC: Line ID search - Discussion - PtII}

\begin{figure}
    \centering
    \includegraphics[width=\linewidth]{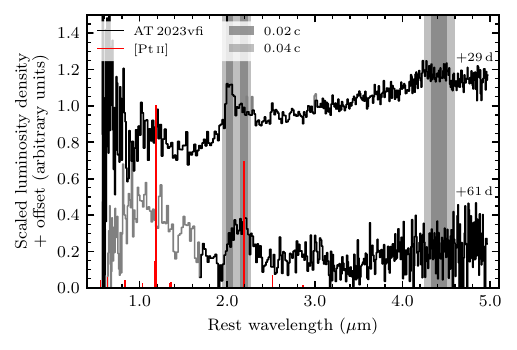}
    \caption{
        Same as Figure~\ref{FIG:TeI+II+III}, but here we present the emission line spectra for [Pt\,\II].
    }
    \label{FIG:PtII}
\end{figure}

We identify two prominent [Pt\,\II] lines here; one at 21883\,\AA, which is coincident with the 2.1874\,\micron\ feature, and the other at 11877\,\AA\ (see Figure~\ref{FIG:PtII} and Table~\ref{TAB:Candidate lines}). Curiously, this 11877\,\AA\ line aligns with a noisy flux excess present in the observational data from +29\,d. While the SNR of the spectrum at these wavelengths is low and thus we do not independently identify it as a prominent emission feature, we note that this [Pt\,\II] 11877\,\AA\ line agrees well with this excess (which also appears to be present in the photometric data; see Figure~\ref{FIG:AT2023vfi +29d spectral fit (NIRSpec)}).

\cite{Gillanders2021} present detailed atomic data for Pt\,\I--\III\ \citep[see also][]{Bromley2023}, which they use to generate synthetic spectra for forbidden emission for each ion. They find that the two strongest lines in their analysis for Pt\,\II\ are also the 11877 and 21883\,\AA\ transitions we shortlist here. Their data include \Aval\ for all transitions, and for the 11877 and 21883\,\AA\ lines they quote \Aval\ of 9.05 and 2.54\,s$^{-1}$, respectively. Considering this additional information, we conclude that our shortlisted 21883\,\AA\ line is expected to be weaker than our estimate here ($\approx 0.3 \times$ as strong). 

From a nucleosynthetic standpoint, one expects Pt to be prevalent in KN ejecta. In the Solar \rpro\ abundance of \cite{Goriely1999} \citep{Prantzos2020}, Pt is the fifth (third) most abundant \rpro\ element for $Z \geq 38$ (see Figure~\ref{FIG:r-process abundances}). In the \YeZeroFive\ composition profile presented by \cite{Gillanders2022_PaperI} (see also Figure~\ref{FIG:r-process abundances}), Pt is actually the most abundant element by mass. Thus, assuming \vfi\ underwent heavy \rpro\ nucleosynthesis, we can reasonably expect Pt to be present, and if it were to produce detectable forbidden emission, we expect it to be attributable predominantly to the 11877 and 21883\,\AA\ lines.

\subsection{Interpreting the $\sim 2.1$\,\micron\ emission as a single feature} \label{SEC: Line ID search - Interpreting the 2.1um emission as a single feature}

Hitherto, we have modelled the observed $\sim 2.1$\,\micron\ feature at +29~days as two superimposed emission components. The simplest explanation for the cause of these two features is the presence of two separate, strong emission transitions, each centred on one of the components. Here we consider whether these two features can be explained by a single transition consistent with the shortlisted [Te\,\III] 21050\,\AA\ line.

In principle, this transition, centred almost exactly between the 2.0218 and 2.1874\,\micron\ features, could explain both emission features if the powering species was abundant in the outermost ejecta material moving directly towards and away from us, either due to physical stratification of elemental abundance or ionisation/recombination effects. Both a blueshifted and redshifted component are required, each offset from the rest wavelength of [Te\,\III] $\lambda 21050$ by $\approx 11800$\,\kms. This corresponds to $0.36 \times v_{\textsc{fwhm}}$ of the broader component. The component luminosities are comparable, but the blueshifted feature possesses approximately half the velocity width (see Table~\ref{TAB:Feature parameters}).

Ejecta inhomogeneities may also play a role, given the asymmetric nature of kilonovae \citep[owing to their multiple distinct ejecta components;  \eg,][]{Kasen2017}. It is possible that the emitting species could be distributed  such that it is disproportionately moving towards (or away from) us. This will lead to observed features that are bulk blueshifted (redshifted) from the rest wavelength of the transition.

Double-peaked emission features have been detected in nebular phase spectra of supernovae. Their structure has been proposed to indicate an aspherical or bipolar explosion, producing a torus, or disc, of emitting material \citep{Maeda2008,Tanaka2009}. In most cases, the red and blue components are symmetric around the rest wavelength of the transition, but in quite a number of cases, a strong blueshifted emission component is observed \citep{Milisavljevic2010,Jerkstrand2015}. This blueshift can be a significant fraction of the $v_{\textsc{fwhm}}$ of the emission feature ($25 - 35$~per~cent in the case of Mg\,{\sc i}] and [O\,{\sc i}]), a similar magnitude to what we would infer for \vfi\ if the $\sim 2.1$\,\micron\ emission is due to the [Te\,\III] $\lambda 21050$ transition.

Invoking some asymmetry seems like a plausible solution to the observed two-component nature of the $\sim 2.1$\,\micron\ feature at +29~days. However, it cannot be due to a viewing-angle effect of an equatorial torus or disc of emitting material, given our viewing angle of \vfi\ is close to on-axis (compared to \eg, \citealt{Tanaka2009}, where they invoke a viewing angle $>50^\circ$ to explain the observed double-peaked structure of the [O\,\I] $\lambda \lambda 6300, 6364$ emission feature in nebular spectra of \SNxx{2008D}). Some other asymmetric ejecta structure must exist that can give rise to a similar observational effect along our polar sightline. This asymmetric structure must also be reconcilable with the blueshifted component being confined to a smaller velocity range than the redshifted component (given the different measured $v_{\textsc{fwhm}}$). Finally, the absence of the blueshifted component at +61~days is not readily explainable by invoking asymmetries, given the redshifted component is still prominent and detectable.

One could also invoke opacity effects to explain the observed feature sub-structure. In the case where the KN ejecta has not yet entered a fully nebular regime, there may still be some residual opacity associated with particular transitions. The emission from any ejecta material moving away from us may be re-processed, and thus we will see  predominantly blueshifted emission. Alternatively, opacity sources such as dust may impact observed feature shape. Newly formed dust in the ejecta could be a source of high opacity, attenuating emission from atomic transitions. This effect could manifest as observed blueshifted features if the dust forms in the inner regions of the ejecta, thus not impacting the emission from the material moving towards us. However, invoking strong opacity to suppress emission from the inner ejecta regions may impact any emission from the ejecta material moving away from us; hence the presence of the redshifted component is not easily explained. And, as was the case for invoking ejecta asymmetries above, the evolution in feature structure from +29 to +61~days cannot be easily explained by opacity effects.

\subsection{Comparison with previous line identification studies}

\subsubsection{\vfi}

\cite{Levan2024} analyse the \jwst\ spectra of \thisGRB\ and propose [Te\,\III] as the source of the emission at $\sim 2.1$\,\micron. Although they empirically model the $\sim 2.1$\,\micron\ flux excess as two blended emission components, they invoke a single component when attributing it to [Te\,\III]; they do not discuss in detail the apparent discrepancy with the rest wavelength of the prominent 21050\,\AA\ line and the observed feature emission. They also find in their modelling that [W\,\III] $\lambda 44322$ and [Se\,\III] $\lambda 45549$ can plausibly produce detectable emission at $\sim 4.5$\,\micron. Here we note that we recover the same proposed identifications for [Te\,\III], [W\,\III] and [Se\,\III],\footnote{We do not highlight [Se\,\III] in Section~\ref{SEC: Line ID search - Discussion} as a species of particular interest as we do not find any reason to favour it over some of our other candidate ions in our analysis (see Section~\ref{SEC: Line ID search - Candidate ions}). However, \cite{Hotokezaka2022} and \cite{Levan2024} highlight [Se\,\III] as a plausible source of late-time emission at $\sim 4.5$\,\micron\ in \gfo\ and \vfi, respectively, based on abundance considerations.} although we expand the list of plausible candidate ions to also include other species.

\cite{Gillanders2023arxiv_GRB230307A} present an independent study of the spectra of \thisGRB, and propose a number of candidate ions. Notably, they do not shortlist [Se\,\III] as a candidate ion for the $\sim 4.4$\,\micron\ emission component, as here we \citep[and][]{Levan2024} have shortlisted in our analyses. The analysis presented here is more robust than that presented by \cite{Gillanders2023arxiv_GRB230307A}. Not only have we re-reduced and carefully flux-calibrated the \jwst\ NIRSpec spectra to obtain the best available version of these data, we have also improved upon the spectral fitting procedure, providing us with more reliable estimates for the continuum and feature emission properties. As such, we have a better estimate of the feature centroids, for which accurate values are of paramount importance in these types of line identification searches. Thus, we argue that the list of proposed line identifications presented here represents a significant improvement over previous searches, and the ions proposed here are those that warrant further prioritised study.

\subsubsection{\gfo}

\cite{Hotokezaka2023} present an analysis of the late-time \xsh\ spectra of \gfo, and show that, assuming the Solar \rpro\ abundance profile for the first \rpro\ peak and above (\ie, $A \geq 88$), one expects the most prominent emission to arise from [Te\,\III] $\lambda 21050$. Their models with Te emission provide a reasonable match to the data, although they do not reproduce the apparent redshifting nature of the $\sim 2.1$\,\micron\ emission across the $+7.4 - 10.4$\,d sequence of \xsh\ spectra.

If the contribution from lighter species is also considered (\ie, $A \geq 69$; again invoking the Solar \rpro\ abundance profile), \cite{Hotokezaka2023} find that strong emission arises from [Kr\,\III] $\lambda 21986$ and [Se\,\IV] $\lambda 22850$. Both of these lines are too red to produce the $\sim 2.1$\,\micron\ emission in \gfo, but here we shortlist the [Kr\,\III] line as a candidate for the 2.1874\,\micron\ feature in \vfi\ (the [Se\,\IV] line is also too red to match the 2.1874\,\micron\ feature).

\cite{Gillanders2023_PaperII} present an analysis of the post-photospheric phase spectra of \gfo, and shortlist a number of species that plausibly produce the $\sim 2.1$\,\micron\ emission. They find the data are best reproduced by invoking two blended emission components (with centroids of 20590, 21350\,\AA), with the apparent redshifting nature of the feature across the late-time spectra readily explained by an evolution in the relative strengths of the two emission components.

They favour [Te\,\III] $\lambda 21050$ as a promising candidate for (one of the components of) the emission at 2.1\,\micron\ in \gfo, but they present alternative species that may also contribute to this emission. We recover a number of their shortlisted ions here as candidates for the emission in \vfi\ ([Pd\,\III], [Ag\,\III], [Te\,\I], [Te\,\III], [Ba\,\II],\footnote{We shortlist [Ba\,\II] here as a candidate ion, but ultimately rule it out following some additional considerations (see Section~\ref{SEC: Line ID search - Discussion - BaII + RaII}). This argument also applies to the observations of \gfo; \ie, we rule out [Ba\,\II] as a candidate ion for \gfo.} [Er\,\II] and [Ir\,\II]).

\subsection{Comparison with \gfo}

\begin{figure}
    \centering
    \includegraphics[width=\linewidth]{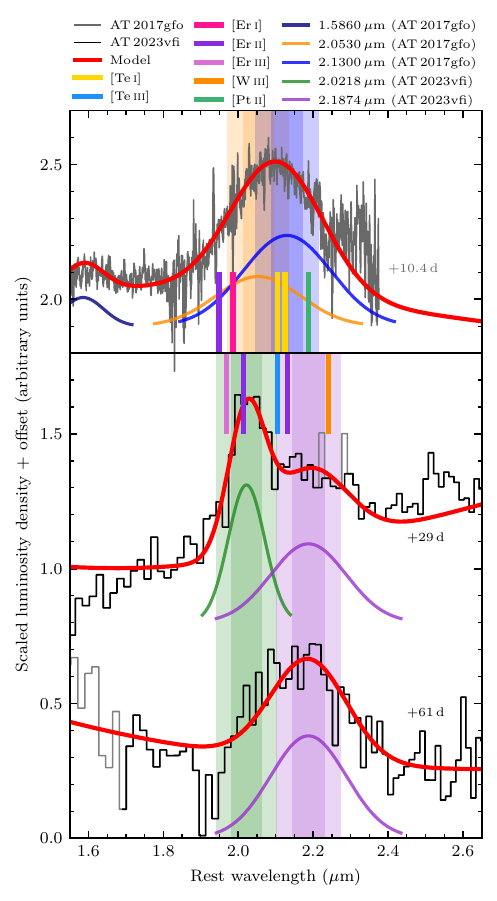}
    \caption{
        Comparison between the +10.4\,d \xsh\ spectrum of \gfo\ (top) and the +29 (middle) and +61\,d (bottom) \jwst\ NIRSpec spectra of \vfi. Our best-fitting models (see Section~\ref{SEC: Spectral fitting}) are overlaid, with the constituent Gaussian emission components for each model also plotted. The 0.02 and 0.04\,c tolerable windows for the 2.0530, 2.1300 (\gfo), 2.0218 and 2.1874\,\micron\ (\vfi) features are also plotted (dark and light shaded bands correspond to the 0.02 and 0.04\,c tolerable windows, respectively; coloured to match the corresponding Gaussian emission component). The shortlisted line identifications highlighted in Section~\ref{SEC: Line ID search - Discussion} are plotted as vertical bars, centred on their rest wavelengths.
    }
    \label{FIG:ATs 2017gfo + 2023vfi line IDs}
\end{figure}

In Figure~\ref{FIG:ATs 2017gfo + 2023vfi line IDs}, we compare the +10.4~day \xsh\ spectrum of \gfo\ with the +29 and +61~day \jwst\ NIRSpec spectra of \vfi. The presence of prominent emission in all spectra at $\sim 2.1$\,\micron\ is clear. While the emission in the +10.4\,d \gfo\ and +29\,d \vfi\ spectra in both cases seems to be composed of two blended emission components, it is more visually apparent in the +29\,d \vfi\ spectrum. Fitting the spectra in a consistent manner (see Section~\ref{SEC: Spectral fitting}) favours similar, but not exactly matching, line centroids for the two components in each system. For \gfo, we constrain the line centroids to be $20530^{+140}_{-190}$ and $21300^{+330}_{-260}$\,\AA, while for \vfi\ we estimate centroids of $20218^{+37}_{-38}$ and $21874 \pm 89$\,\AA\ (see Table~\ref{TAB:Feature parameters}).

Considering the tolerable ranges we invoked for our line identification searches (0.02 and 0.04\,c), there is significant overlap between the shortlisted lines for each system (see Figure~\ref{FIG:ATs 2017gfo + 2023vfi line IDs}). It seems natural to expect such prominent emission at very similar wavelengths in both systems to be attributed to some common species present in the ejecta material of both systems. If so, the feature differences between these two events would allow us to compare the ejecta morphology and composition of each system, since feature strength, width, and offset from the rest wavelength of the powering transition can all be used to infer ejecta properties (including \eg, composition, expansion velocity, ionisation, excitation etc.). However, common candidate transitions in both events need to be confirmed before such inferences are made. In the case of the blended $\sim 2.1$\,\micron\ emission discussed here, the small deviation in feature centroids between \gfo\ and \vfi\ can be explained by \eg, different ejecta distribution, different opacity effects, or the blending of contributing (but not dominating) emission from other species.

Further study is needed to more fully investigate the effects of line blending and opacity to determine whether the blended emission at $\sim 2.1$\,\micron\ in both \gfo\ and \vfi\ can be explained by one (or multiple) common source(s). For now, it seems likely that [Te\,\III] contributes to (or dominates) this emission in both systems \citep[as proposed by][]{Hotokezaka2023, Gillanders2023arxiv_GRB230307A, Gillanders2023_PaperII, Levan2024}.

\section{Conclusions} \label{SEC: Conclusions}

\vfi, the kilonova associated with \thisGRB, is just the second kilonova to be observed spectroscopically. It also marks the first \jwst\ observations of a kilonova, with two NIRSpec spectra obtained at phases of +29 and +61~days post-GRB. In this paper, we took an empirical approach to fitting the observed spectra. After re-reducing the spectra and calibrating to contemporaneous \jwst\ NIRCam photometry, we estimated the continuum properties and fit the observed flux excesses at $\sim 2.1$ and 4.4\,\micron, which we interpret as emission features. The X-ray to MIR continua can be well-matched by the combination of a power law and a blackbody,  representative of the contributions of the non-thermal afterglow and the rising red continuum at wavelengths $\gtrsim 2$\,\micron, respectively. We also fit the prominent flux excesses as Gaussian emission features, and find the +29~day spectrum is best fit invoking three emission components at $20218_{-38}^{+37}$,  $21874 \pm 89$ and $44168_{-152}^{+153}$\,\AA. The $\sim 2.1$\,\micron\ flux excess at +61~days is well-reproduced by a single component with the same centroid and width as the $21874 \pm 89$\,\AA\ feature invoked at +29~days. The Gaussian features have velocity widths of \mbox{$0.057 \lesssim v_{\textsc{fwhm}} ({\rm c}) \lesssim 0.110$}.

We undertook a complete line identification search, with all publicly available atomic data, considering forbidden transitions for the lowest four ionisation stages (neutral to triply ionised) of the elements with atomic number, $Z = 30 - 92$. We shortlisted a number of ions that have transitions coincident with the observed emission features. This list includes the `light' ($Z \leq 56$) \rpro\ species   Se\,\III, Kr\,\III, Nb\,\IV, Pd\,\III, Ag\,\III, In\,\I, Te\,\I--\III, and Ba\,\II, the lanthanide ($57 \leq Z \leq 71$) species Ce\,\IV, Sm\,\IV, Ho\,\IV, and Er\,\I--\III, and the `heavy' ($Z \geq 72$) \rpro\ species Hf\,\I, W\,\III, Os\,\I, Ir\,\II, Pt\,\II, Ra\,\II, and Ac\,\I.

Considering all possible transitions and \rpro\ element abundance patterns, we conclude that the astrophysically observed [Te\,\III] 21050\,\AA\ line remains the best candidate to explain the observed emission profiles at 2.0218 and 2.1874\,\micron\ in \vfi. This is in agreement with previous work \citep[][]{Gillanders2023arxiv_GRB230307A, Levan2024}, but we highlight that the rest wavelength of [Te\,\III] $\lambda 21050$ does not match the centroid of either of the components (it is offset from each by $v \simeq 0.04$\,c); instead, it resides exactly at the average of the two. Strong opacity effects, bulk velocity structure, or aspherical geometries are required to reconcile the observed feature positions with the rest wavelength of this line. Such feature structure, and blueshifts,  have been observed in the nebular spectra of type Ib/c supernovae. Te\,\III\ has another expected transition at 29290\,\AA\ that we do not detect. However, we estimate it should be around 10~per~cent (20~per~cent) of the luminosity of the [Te\,\III] $\lambda 21050$ line at $T = 3000$\,K ($T = 5000$\,K); we show that the noise in the NIRSpec data around this wavelength precludes a detection. W\,\III\ can plausibly match multiple emission features; we find good agreement between its 22416, 44322 and 45352\,\AA\ transitions and the 2.1874 and 4.4168\,\micron\ emission features. These ions are of particular interest since they are both even-$Z$ elements that lie at the top of the second (Te), or close to the third (W), \rpro\ abundance peaks. 

We quantitatively compared the $2.1$\,\micron\ features of \vfi\ and \gfo. The wavelength coincidence of these features has been noted by \cite{Levan2024} and \cite{Gillanders2023arxiv_GRB230307A}, who invoked [Te\,\III] $\lambda 21050$ as the explanation. Similar to our analysis of the \vfi\ feature, we find that two components are preferred to fit the feature in \gfo. However, the measured line centres are neither coincident between the two kilonovae, nor consistent with a single velocity offset. These inconsistencies prevent us from confirming that the [Te\,\III] 21050\,\AA\ transition is definitely the cause of the emission in both objects. However, we consider it to be the most plausible interpretation. The alternative scenario is that one (or both) of the components of the $\sim 2.1$\,\micron\ emission feature are instead powered by transitions between some low-lying levels of one (or multiple) as-yet unidentified \rpro\ species.

Our analysis suffers from two limitations, the first of which is incomplete atomic data, specifically: (i) the incomplete line lists for the heavy \rpro\ elements, which are typically lacking in forbidden transition wavelengths and intrinsic line strengths, and (ii) the non-existence of collisional information for most of the \rpro\ species. Non-LTE effects are crucial to fully understand late-time properties of kilonova ejecta, but without these data, here we are limited to using LTE approximations for our level population calculations. Second, our approach does not account for complex radiative transfer and non-LTE effects. A detailed study, utilising quantitative kilonova modelling codes \citep[such as \eg, \textsc{artis} or \textsc{sumo};][]{Kromer-ARTIS-2009, Jerkstrand-SUMO-2011, Shingles2020, Shingles2023, Pognan2022, Pognan2022_nlte, Pognan2023} should be undertaken, to fully analyse these \jwst\ spectra of \vfi. While we get good agreement here to the continua of the observed spectra invoking a thermal blackbody component, it remains unclear whether this is physically reasonable at such late phases, or whether the data are actually produced via some other mechanism; \eg, fluorescence.

These \jwst\ data of \vfi\ are vital for advancing our understanding of the late-phase evolution of KNe, especially in the NIR. Hence we have taken care to perform a detailed extraction and flux calibration of these data, which we make public for use in future studies.  \jwst\ will play a large part in advancing our understanding of a growing set of kilonova events.

\begin{table*}
    \renewcommand*{\arraystretch}{1.2}
    \centering
    \caption{
        List of candidate transitions for the inferred emission features at 2.0218, 2.1874 and 4.4168\,\micron. For each transition, we provide the species, wavelength (in vacuum and in air), line strength, and the level information (configuration, term, quantum $J$ number and energy) for both the upper and lower levels.
    }
    \begin{threeparttable}
        \centering
        \begin{tabular}{lcccrlcccrlcc}
        \toprule
        
        Species    &$\lambda_{\rm vac}$ (\AA)    &$\lambda_{\rm air}$ (\AA)    &Line strength$^{\rm a}$    &\multicolumn{4}{c}{Lower level}    &    &\multicolumn{4}{c}{Upper level}    \\
        \cline{5-8}
        \cline{10-13}
            &    &      &($T = 3000$\,K)    &Configuration    &Term    &$J$    &Energy (eV)    &    &Configuration    &Term    &$J$    &Energy (eV)    \\

        \midrule
        \multicolumn{13}{l}{2.0218\,\micron\ feature (0.02\,c)} \\
        \midrule

        $[^{56}$Ba\,\II] 		 &20518 		 &20512       &0.84 	 &6s 		             &$^2$S 		                         &\sfrac{1}{2} 		 &0.000 		 & 		 &5d 		             &$^2$D                                  &\sfrac{3}{2} 		 &0.604 			 \\ 
        $[^{68}$Er\,\I] 		 &19860 		 &19855       &1.0 		 &4f$^{12}$6s$^2$ 		 &$^3$H 		                         &6 		         &0.000 		 & 		 &4f$^{12}$6s$^2$ 		 &$^3$F		                             &4          		 &0.624 			 \\ 
        $[^{68}$Er\,\II] 		 &20148 		 &20143       &0.68      &$-$              		 &$-$      		                         &\sfrac{11}{2} 	 &0.055 		 & 		 &$-$              		 &$-$  		                             &\sfrac{7}{2} 		 &0.670 			 \\ 
        $[^{88}$Ra\,\II] 		 &20587 		 &20581       &1.0 		 &7p 		             &$^2$P$^{\textrm{\textsc{o}}}$ 		 &\sfrac{1}{2} 		 &2.647 		 & 		 &7p 		             &$^2$P$^{\textrm{\textsc{o}}}$ 		 &\sfrac{3}{2} 		 &3.249 			 \\ 

        \midrule
        \multicolumn{13}{l}{2.0218\,\micron\ feature (0.04\,c)} \\
        \midrule

        $[^{67}$Ho\,\IV] 		 &19806 		 &19800         &1.0 		 &4f$^{10}$ 		 &$^5$I 		 &8              		 &0.000 		 & 		 &4f$^{10}$ 		 &$^5$I 		 &7            		 &0.626 			 \\ 
        $[^{68}$Er\,\II] 		 &19483 		 &19478         &1.0 		 &$-$ 		         &$-$ 		     &\sfrac{13}{2}  		 &0.000 		 & 		 &$-$          		 &$-$     		 &\sfrac{9}{2} 		 &0.636 			 \\ 
        $[^{68}$Er\,\III] 		 &19678 		 &19673         &1.0 		 &4f$^{12}$ 		 &$^3$H 		 &6             		 &0.000 		 & 		 &4f$^{12}$ 		 &$^3$F 		 &4            		 &0.630 			 \\ 
        $[^{76}$Os\,\I] 		 &19440 		 &19435         &1.0 		 &5d$^6$6s$^2$ 		 &$^5$D 		 &4                		 &0.000 		 & 		 &5d$^7$($^4$F)6s 	 &$^5$F 		 &5            		 &0.638 			 \\ 
        $[^{77}$Ir\,\II] 		 &20886 		 &20880         &1.0 		 &5d$^7$($^4$F)6s 	 &$^5$F 		 &5             		 &0.000 		 & 		 &5d$^7$($^4$F)6s 	 &$^5$F 		 &4           		 &0.594 			 \\ 

        \midrule
        \multicolumn{13}{l}{2.1874\,\micron\ feature (0.02\,c)} \\
        \midrule

        $[^{36}$Kr\,\III] 		 &21986 		&21980       &1.0 		 &4s$^2$4p$^4$ 		 &$^3$P 		 &2         		 &0.000 		 & 		 &4s$^2$4p$^4$ 		 &$^3$P 		 &1         		 &0.564 			 \\ 
        $[^{47}$Ag\,\III] 		 &21696 		&21690       &1.0 		 &4d$^9$ 		     &$^2$D 		 &\sfrac{5}{2} 		 &0.000 		 & 		 &4d$^9$ 	    	 &$^2$D 		 &\sfrac{3}{2} 		 &0.571 			 \\ 
        $[^{72}$Hf\,\I] 		 &21893 		&21887       &0.86 		 &5d$^2$6s$^2$ 		 &$^3$F 		 &2          		 &0.000 		 & 		 &5d$^2$6s$^2$ 		 &$^3$F 		 &4         		 &0.566 			 \\ 
        $[^{78}$Pt\,\II] 		 &21883 		&21877       &0.69 		 &5d$^8$6s 		     &$^4$F 		 &\sfrac{9}{2} 		 &0.593 		 & 		 &5d$^8$6s 		     &$^4$F 		 &\sfrac{7}{2} 		 &1.160 			 \\ 

        \midrule
        \multicolumn{13}{l}{2.1874\,\micron\ feature (0.04\,c)} \\
        \midrule

        $[^{46}$Pd\,\III] 		 &21338 	&21332	 &0.52 		 &4d$^8$ 		 &$^3$F 		 &4                		 &0.000 		 & 		 &4d$^8$ 		     &$^3$F 		 &2           		 &0.581 			 \\ 
        $[^{52}$Te\,\I] 		 &21049 	&21044	 &1.0 		 &5p$^4$ 		 &$^3$P 		 &2               		 &0.000 		 & 		 &5p$^4$ 		     &$^3$P 		 &1           		 &0.589 			 \\ 
        $[^{52}$Te\,\I] 		 &21247 	&21241	 &0.34 		 &5p$^4$ 		 &$^3$P 		 &2              		 &0.000 		 & 		 &5p$^4$ 		     &$^3$P 		 &0           		 &0.584 			 \\ 
        $[^{52}$Te\,\III] 		 &21050 	&21044	 &1.0 		 &5s$^2$5p$^2$	 &$^3$P 		 &0                      &0.000 		 & 		 &5s$^2$5p$^2$ 		 &$^3$P 		 &1           		 &0.589 			 \\ 
        $[^{68}$Er\,\II] 		 &21312 	&21306	 &0.91 		 &$-$ 	      	 &$-$ 	    	 &\sfrac{11}{2} 		 &0.055 		 & 		 &$-$ 	     	     &$-$ 	    	 &\sfrac{9}{2} 		 &0.636 			 \\ 
        $[^{74}$W\,\III] 		 &22416 	&22409	 &1.0 		 &5d$^4$ 		 &$^5$D 		 &0               		 &0.000 		 & 		 &5d$^4$ 		     &$^5$D 		 &2          		 &0.553 			 \\ 

        \midrule
        \multicolumn{13}{l}{4.4168\,\micron\ feature (0.02\,c)} \\
        \midrule

        $[^{58}$Ce\,\IV] 		 &44391 	&44379           &1.0 		 &5p$^6$4f 		 &$^2$F$^{\textrm{\textsc{o}}}$ 	 &\sfrac{5}{2} 		 &0.000 		 & 		 &5p$^6$4f 		 &$^2$F$^{\textrm{\textsc{o}}}$		 &\sfrac{7}{2} 		 &0.279 			 \\ 
        $[^{62}$Sm\,\IV] 		 &43656 	&43645           &1.0 		 &4f$^5$ 		 &$^6$H$^{\textrm{\textsc{o}}}$ 	 &\sfrac{5}{2} 		 &0.000 		 & 		 &4f$^5$ 		 &$^6$H$^{\textrm{\textsc{o}}}$ 	 &\sfrac{9}{2} 		 &0.284 			 \\ 
        $[^{74}$W\,\III] 		 &44322 	&44310           &0.87 		 &5d$^4$ 		 &$^5$D 		                     &0         		 &0.000 		 & 		 &5d$^4$ 		 &$^5$D                        		 &1           		 &0.280 			 \\ 
        $[^{89}$Ac\,\I] 		 &44814 	&44802           &1.0 		 &6d7s$^2$ 		 &$^2$D 		                     &\sfrac{3}{2} 		 &0.000 		 & 		 &6d7s$^2$ 		 &$^2$D                      		 &\sfrac{5}{2} 		 &0.277 			 \\ 
        
        \midrule
        \multicolumn{13}{l}{4.4168\,\micron\ feature (0.04\,c)} \\
        \midrule

        $[^{34}$Se\,\III] 		 &45549 	&45537	 &0.56 		 &4s$^2$4p$^2$ 		 &$^3$P 		                         &1          		 &0.216 		 & 		 &4s$^2$4p$^2$ 		 &$^3$P                          		 &2            		 &0.488 			 \\ 
        $[^{41}$Nb\,\IV] 		 &42637 	&42625	 &1.0 		 &4d$^2$      		 &$^3$F 		                         &2         		 &0.000 		 & 		 &4d$^2$     		 &$^3$F                           		 &4         		 &0.291 			 \\ 
        $[^{49}$In\,\I] 		 &45196 	&45183	 &1.0 		 &5s$^2$5p           &$^2$P$^{\textrm{\textsc{o}}}$ 		 &\sfrac{1}{2} 		 &0.000 		 & 		 &5s$^2$5p   		 &$^2$P$^{\textrm{\textsc{o}}}$ 		 &\sfrac{3}{2} 		 &0.274 			 \\ 
        $[^{52}$Te\,\II] 		 &45466 	&45453	 &1.0 		 &5s$^2$5p$^3$ 		 &$^2$D$^{\textrm{\textsc{o}}}$ 		 &\sfrac{3}{2} 		 &1.267 		 & 		 &5s$^2$5p$^3$ 		 &$^2$D$^{\textrm{\textsc{o}}}$ 		 &\sfrac{5}{2} 		 &1.540 			 \\ 
        $[^{72}$Hf\,\I] 		 &42433 	&42421	 &1.0 		 &5d$^2$6s$^2$ 		 &$^3$F 		                         &2         		 &0.000 		 & 		 &5d$^2$6s$^2$ 		 &$^3$F                         		 &3         		 &0.292 			 \\ 
        $[^{72}$Hf\,\I] 		 &45229 	&45217	 &0.42 		 &5d$^2$6s$^2$ 		 &$^3$F 		                         &3            		 &0.292 		 & 		 &5d$^2$6s$^2$ 		 &$^3$F                            		 &4          		 &0.566 			 \\ 
        $[^{74}$W\,\III] 		 &45352 	&45339	 &0.49 		 &5d$^4$      		 &$^5$D 		                         &1            		 &0.280 		 & 		 &5d$^4$ 	      	 &$^5$D                         		 &2 	           	 &0.553 			 \\ 
        
        \bottomrule
        \end{tabular}
        \begin{tablenotes}
            \footnotesize
            \item \textbf{Note.} $\lambda_{\rm air}$ values have been computed assuming the standard vacuum-to-air conversion from VALD3 \citep[see][]{Birch1994, Morton2000, VALD3}.
            \item[a] Line strength represents the normalised line luminosity values ($L_{\rm em}$) we computed for each transition (see Section~\ref{SEC: Line ID search - Methodology} for details).
            \item[o] Denotes an odd parity.
        \end{tablenotes}
    \end{threeparttable}
    \label{TAB:Candidate lines}
\end{table*}

\section*{Acknowledgements}

We thank the anonymous reviewer for detailed comments that improved the final manuscript. We thank Alex~J.~Cameron, Andrew~J.~Levan and Lauren~Rhodes for discussions regarding the reduction and interpretation of the \jwst\ NIRSpec spectra, Stuart~A.~Sim for discussions surrounding our line identification search, Lauren Rhodes and Shubham Srivastav for assistance with data visualisation, and the \jwst\ help desk for assisting with data interpretation. S.~J.~Smartt acknowledges funding from STFC Grants ST/Y001605/1, ST/X006506/1 and ST/T000198/1, and a Royal Society Research Professorship. 

\section*{Data Availability}

Our reductions of the \jwst\ NIRSpec spectra of \vfi\ presented in this paper are publicly available at \specURL. The \xsh\ spectra of \gfo\ are publicly available at \url{www.engrave-eso.org/AT2017gfo-Data-Release} and \url{https://wiserep.weizmann.ac.il}. The photometry utilised in this paper has previously been published (see \citealt{Levan2024} for the \jwst\ data, and \citealt{Yang2024} for the \textit{XMM-Newton} and \textit{Chandra} data).

\bibliographystyle{mnras}
\bibliography{References}


\appendix

\section{\vfi\ extended figures}

Here we present the best-fitting models for the +29 and +61~day \jwst\ NIRSpec spectra of \vfi\ extended to shorter wavelengths, to illustrate model agreement with the inter/extra-polated X-ray data (see Section~\ref{SEC: Spectral fitting - AT2023vfi}).

\begin{figure*}
    \centering
    \includegraphics[width=0.8\linewidth]{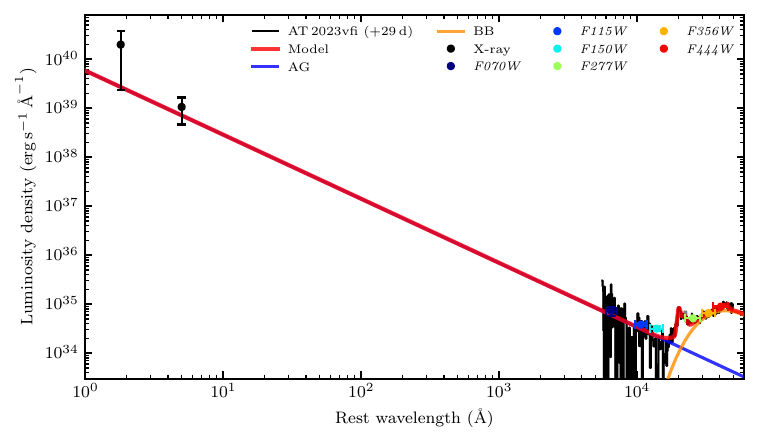}
    \caption{
        Same as Figure~\ref{FIG:AT2023vfi +29d spectral fit (NIRSpec)}, but plotted in log space to accommodate the inclusion of our interpolated X-ray data (black circles; these plotted errors do not include any estimate for the uncertainty associated with data interpolation).
    }
    \label{FIG:Appendix - +29d spectral fit}
\end{figure*}

\begin{figure*}
    \centering
    \includegraphics[width=0.8\linewidth]{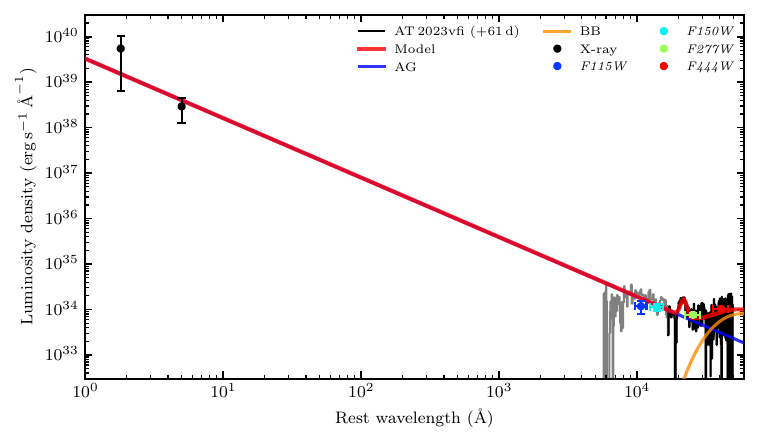}
    \caption{
        Same as Figure~\ref{FIG:AT2023vfi +61d spectral fit (NIRSpec)}, but plotted in log space to accommodate the inclusion of our extrapolated X-ray data (black circles; these plotted errors do not include any estimate for the uncertainty associated with data extrapolation).
    }
    \label{FIG:Appendix - +61d spectral fit}
\end{figure*}

\section{\gfo\ spectral fitting}

Here we present our best-fitting models for the $+7.4 - 10.4$~day \xsh\ spectra of \gfo, as discussed in the main text (see Section~\ref{SEC: Spectral fitting - AT2017gfo}).

\begin{figure*}
    \centering
    \includegraphics[width=0.8\linewidth]{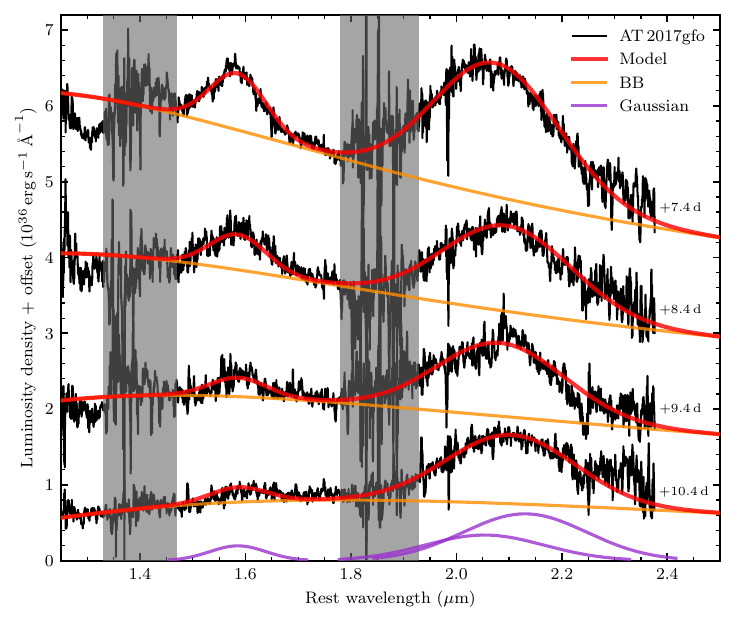}
    \caption{
        Sequence of late-time ($+7.4 - 10.4$\,d) \xsh\ spectra of \gfo\ (black) compared with our best-fitting models (red), offset (by $3.2 \times 10^{36}$, $2.2 \times 10^{36}$ and $10^{36}$\,erg\,s$^{-1}$\,\AA$^{-1}$ for the +7.4, +8.4 and +9.4\,d spectra, respectively) for clarity. The observed spectra are annotated with their phase relative to GW trigger. The continuum BB components for each model are plotted (orange), as are the individual Gaussian components for the +10.4\,d model (purple). These Gaussian components have identical centroids and widths across all phases. The vertical grey bands correspond to strong telluric regions.
    }
    \label{FIG:AT2017gfo spectral fits}
\end{figure*}

\section{Additional line ID plots}

In this section we present the figures for the candidate ions shortlisted in Section~\ref{SEC: Line ID search - Candidate ions} that were not presented in the main text.

We also present sequences of synthetic spectra for a number of ions, as noted in the main text ([Pd\,\III], [Sm\,\IV], [W\,\III] and [Os\,\I]; see Section~\ref{SEC: Line ID search - Candidate ions}), illustrating how they vary across our range of explored $T$ values ($T \in [1000, 3000, 5000]$\,K).

\begin{figure}
    \centering
    \includegraphics[width=\linewidth]{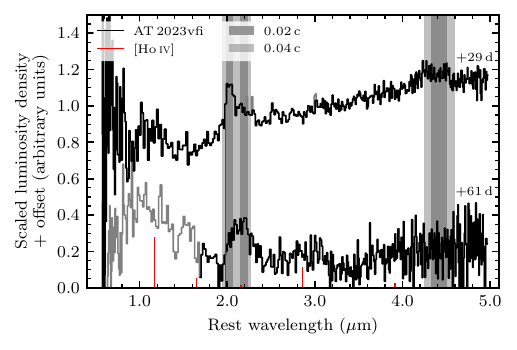}
    \caption{
        Same as Figure~\ref{FIG:TeI+II+III}, but here we present the emission line spectra for [Ho\,\IV].
    }
    \label{FIG:HoIV}
\end{figure}

\begin{figure}
    \centering
    \includegraphics[width=\linewidth]{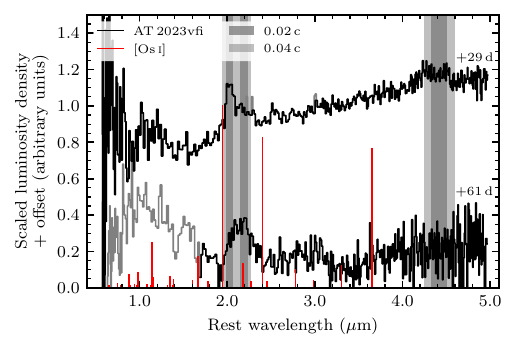}
    \caption{
        Same as Figure~\ref{FIG:TeI+II+III}, but here we present the emission line spectra for [Os\,\I].
    }
    \label{FIG:OsI}
\end{figure}

\begin{figure}
    \centering
    \includegraphics[width=\linewidth]{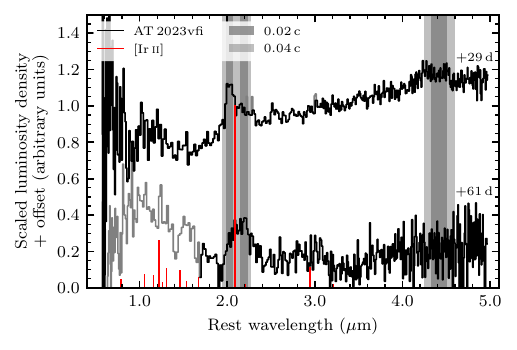}
    \caption{
        Same as Figure~\ref{FIG:TeI+II+III}, but here we present the emission line spectra for [Ir\,\II].
    }
    \label{FIG:IrII}
\end{figure}

\begin{figure}
    \centering
    \includegraphics[width=\linewidth]{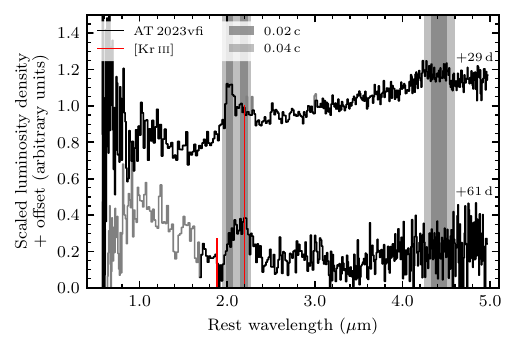}
    \caption{
        Same as Figure~\ref{FIG:TeI+II+III}, but here we present the emission line spectra for [Kr\,\III].
    }
    \label{FIG:KrIII}
\end{figure}

\begin{figure}
    \centering
    \includegraphics[width=\linewidth]{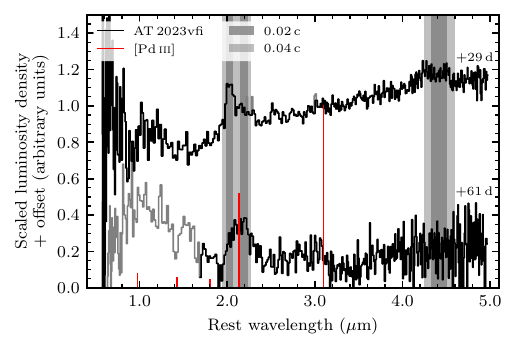}
    \caption{
        Same as Figure~\ref{FIG:TeI+II+III}, but here we present the emission line spectra for [Pd\,\III].
    }
    \label{FIG:PdIII}
\end{figure}

\begin{figure}
    \centering
    \includegraphics[width=\linewidth]{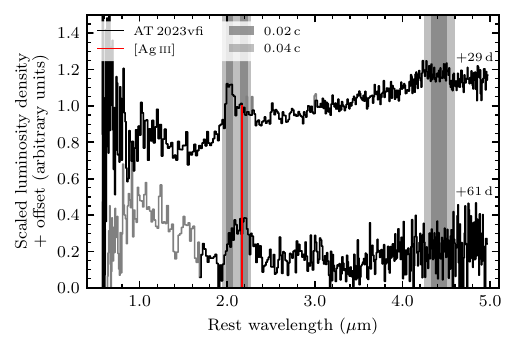}
    \caption{
        Same as Figure~\ref{FIG:TeI+II+III}, but here we present the emission line spectra for [Ag\,\III].
    }
    \label{FIG:AgIII}
\end{figure}

\begin{figure}
    \centering
    \includegraphics[width=\linewidth]{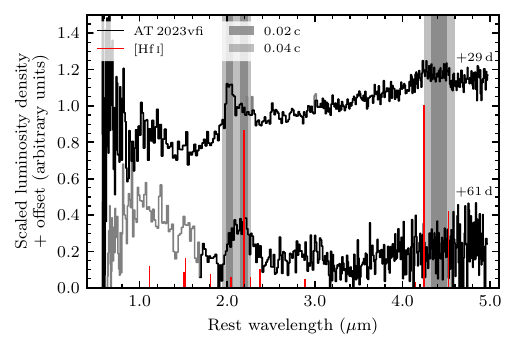}
    \caption{
        Same as Figure~\ref{FIG:TeI+II+III}, but here we present the emission line spectra for [Hf\,\I].
    }
    \label{FIG:HfI}
\end{figure}

\begin{figure}
    \centering
    \includegraphics[width=\linewidth]{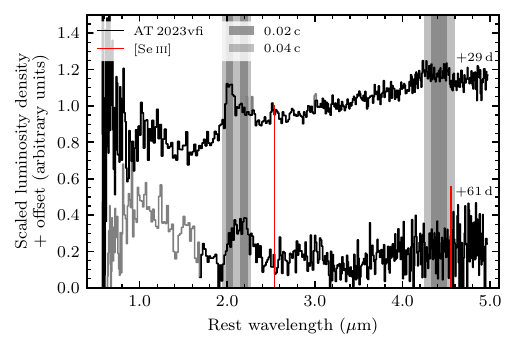}
    \caption{
        Same as Figure~\ref{FIG:TeI+II+III}, but here we present the emission line spectra for [Se\,\III].
    }
    \label{FIG:SeIII}
\end{figure}

\begin{figure}
    \centering
    \includegraphics[width=\linewidth]{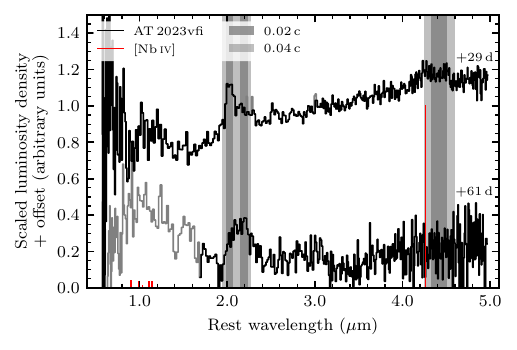}
    \caption{
        Same as Figure~\ref{FIG:TeI+II+III}, but here we present the emission line spectra for [Nb\,\IV].
    }
    \label{FIG:NbIV}
\end{figure}

\begin{figure}
    \centering
    \includegraphics[width=\linewidth]{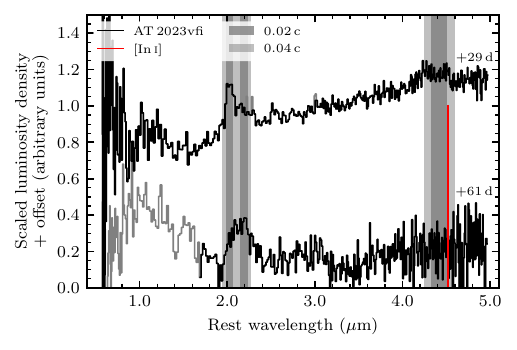}
    \caption{
        Same as Figure~\ref{FIG:TeI+II+III}, but here we present the emission line spectra for [In\,\I].
    }
    \label{FIG:InI}
\end{figure}

\begin{figure}
    \centering
    \includegraphics[width=\linewidth]{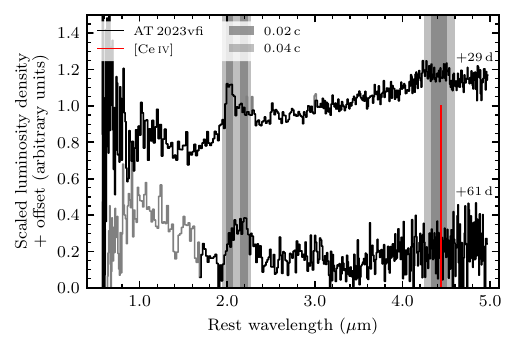}
    \caption{
        Same as Figure~\ref{FIG:TeI+II+III}, but here we present the emission line spectra for [Ce\,\IV].
    }
    \label{FIG:CeIV}
\end{figure}

\begin{figure}
    \centering
    \includegraphics[width=\linewidth]{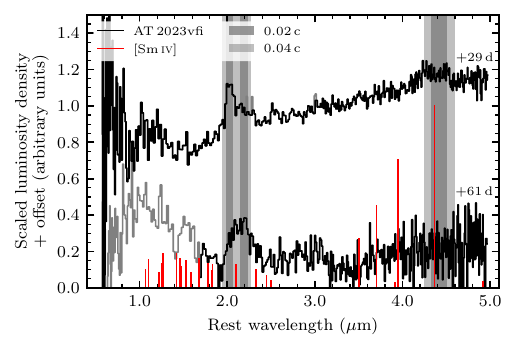}
    \caption{
        Same as Figure~\ref{FIG:TeI+II+III}, but here we present the emission line spectra for [Sm\,\IV].
    }
    \label{FIG:SmIV}
\end{figure}

\begin{figure}
    \centering
    \includegraphics[width=\linewidth]{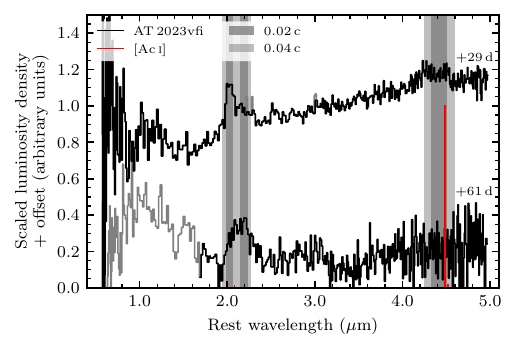}
    \caption{
        Same as Figure~\ref{FIG:TeI+II+III}, but here we present the emission line spectra for [Ac\,\I].
    }
    \label{FIG:AcI}
\end{figure}

\begin{figure}
    \centering
    \subfigure{\includegraphics[width=\linewidth]{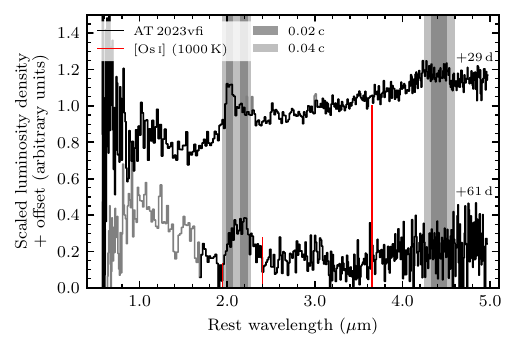}}
    \subfigure{\includegraphics[width=\linewidth]{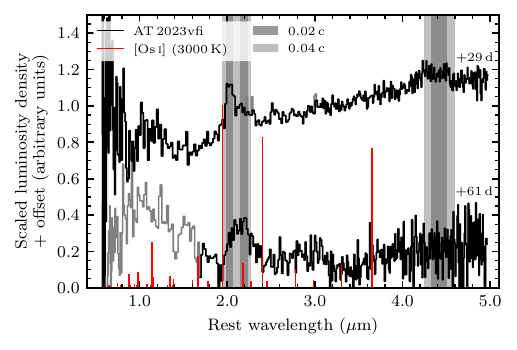}}
    \subfigure{\includegraphics[width=\linewidth]{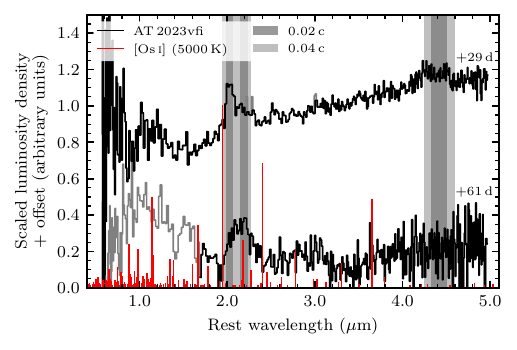}}
    \caption{
        Same as Figure~\ref{FIG:TeI+II+III}, but here we instead present the emission lines for a single species across the range of temperature values probed in our analysis. In this case, we present the relative strengths of the emission transitions of [Os\,\I] at $T = 1000$\,K (top), $T = 3000$\,K (middle), and \mbox{$T = 5000$\,K} (bottom).
    }
    \label{FIG:Candidate ions - Temp evolution - OsI}
\end{figure}

\begin{figure}
    \centering
    \subfigure{\includegraphics[width=\linewidth]{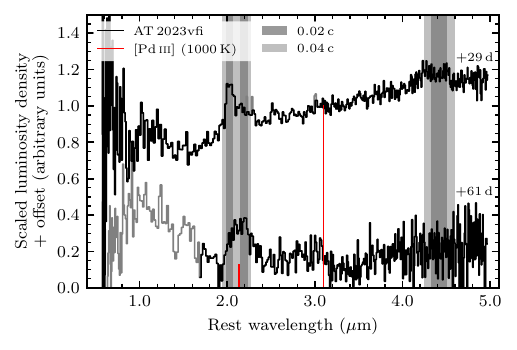}}
    \subfigure{\includegraphics[width=\linewidth]{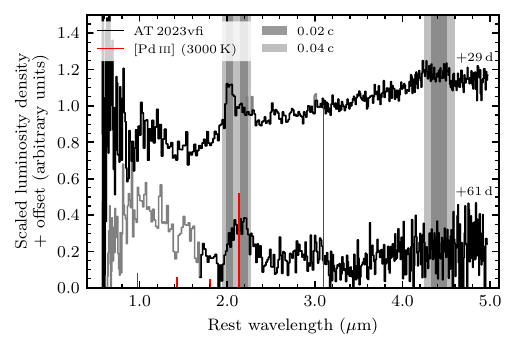}}
    \subfigure{\includegraphics[width=\linewidth]{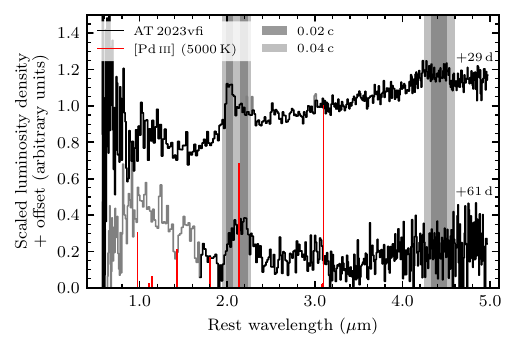}}
    \caption{
        Same as Figure~\ref{FIG:Candidate ions - Temp evolution - OsI}, but in this case, we present the line information for [Pd\,\III].
    }
    \label{FIG:Candidate ions - Temp evolution - PdIII}
\end{figure}

\begin{figure}
    \centering
    \subfigure{\includegraphics[width=\linewidth]{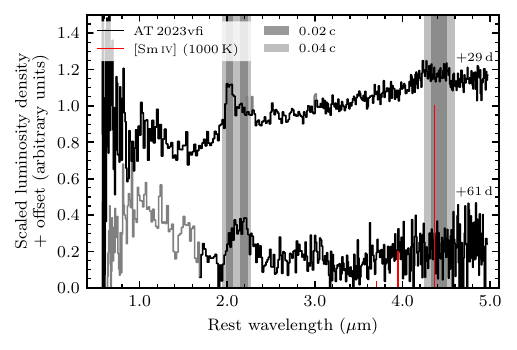}}
    \subfigure{\includegraphics[width=\linewidth]{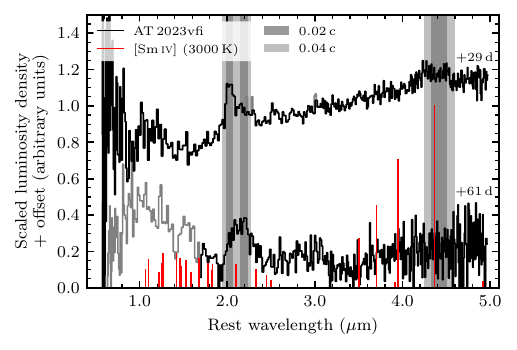}}
    \subfigure{\includegraphics[width=\linewidth]{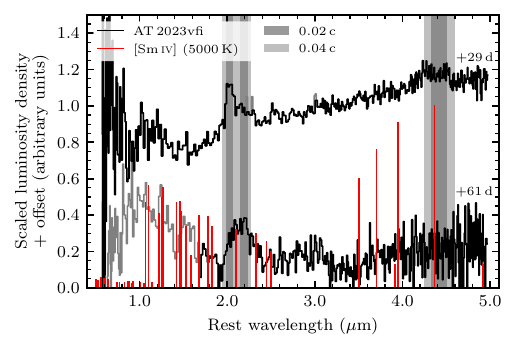}}
    \caption{
        Same as Figure~\ref{FIG:Candidate ions - Temp evolution - OsI}, but in this case, we present the line information for [Sm\,\IV].
    }
    \label{FIG:Candidate ions - Temp evolution - SmIV}
\end{figure}

\begin{figure}
    \centering
    \subfigure{\includegraphics[width=\linewidth]{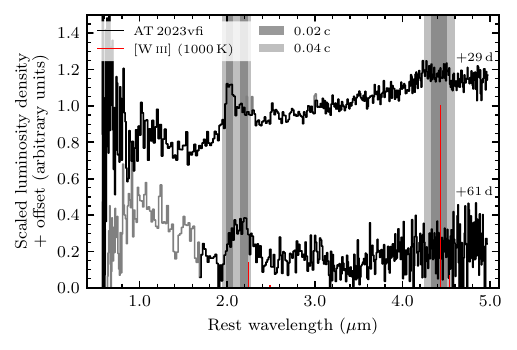}}
    \subfigure{\includegraphics[width=\linewidth]{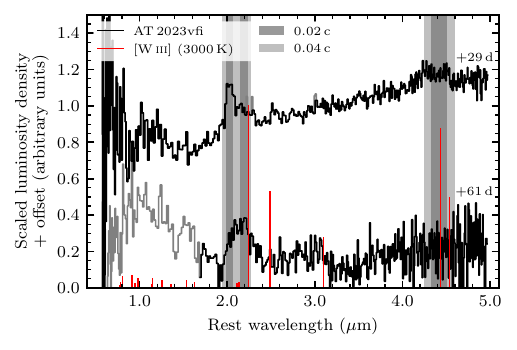}}
    \subfigure{\includegraphics[width=\linewidth]{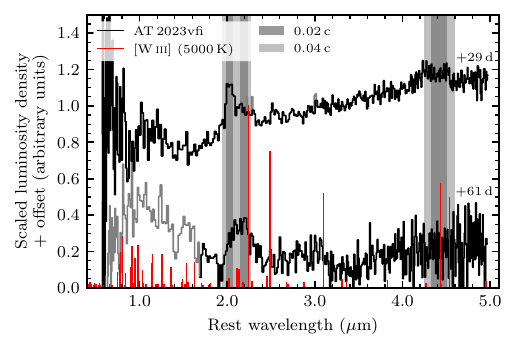}}
    \caption{
        Same as Figure~\ref{FIG:Candidate ions - Temp evolution - OsI}, but in this case, we present the line information for [W\,\III].
    }
    \label{FIG:Candidate ions - Temp evolution - WIII}
\end{figure}


\bsp	
\label{lastpage}
\end{document}